\documentclass[aps, prd,twocolumn,superscriptaddress,nofootinbib,preprintnumbers]{revtex4-1}
\usepackage[utf8]{inputenc}
\usepackage[english]{babel}
\usepackage{natbib}
\usepackage{graphicx}
\usepackage{amsmath,amssymb}
\usepackage{hyperref}
\usepackage{braket}
\usepackage{float}
\usepackage[dvipsnames]{xcolor}
\usepackage{soul}
\usepackage{rotating}
\usepackage{multirow}
\usepackage{mathtools}
\usepackage{makecell}

\usepackage[caption = false]{subfig}
\usepackage{txfonts}
\usepackage[margin=0.75in]{geometry}
\usepackage{tabularx}
\let\oldequation\equation
\let\oldendequation\endequation
\renewenvironment{equation}
  {\linenomathNonumbers\oldequation}
  {\oldendequation\endlinenomath}

\usepackage{hyperref}
\usepackage[T1]{fontenc}
\usepackage{ae,aecompl}

\usepackage{graphicx}	
\usepackage{amsmath}	
\usepackage{adjustbox}
\usepackage{multirow}

\newcommand{\avg}[1]{\langle #1 \rangle}

\newcommand{\vestt}{\mbox{\boldmath $e_{\rm *}$}}

\newcommand{\vest}{\mbox{\boldmath $e$}}

\newcommand{\vecg}{\mbox{\boldmath $\gamma$}}

\newcommand{\vecgest}{\mbox{\boldmath $\gamma^{\mathrm{est}}$}}

\newcommand{\imagunit}{\ensuremath{\textrm{i}\mkern1mu}}

\DeclareMathOperator{\PH}{PH}
\DeclareMathOperator{\Avg}{Avg}
\DeclareMathOperator{\WPHG}{WPHG}
\DeclareMathOperator{\ST}{ST}
\DeclareMathOperator{\Rot}{Rot}
\DeclareMathOperator{\CDF}{CDF}

\definecolor{purple}{RGB}{150,0,200}

\newcommand{\healpix}[0]{\textsc{HEALPix}}

\usepackage{lineno}

\begin{document}		

\title[DES Y3 moments, wavelet phase harmonics, and scattering transform]{Dark Energy Survey Year 3 results: simulation-based cosmological inference with wavelet harmonics, scattering transforms, and moments of weak lensing mass maps I -- validation on simulations}

\label{firstpage}





\begin{abstract}
Beyond-two-point statistics contain additional information on cosmological as well as astrophysical and observational (systematics) parameters. In this methodology paper we provide an end-to-end simulation-based analysis of a set of Gaussian and non-Gaussian weak lensing statistics using detailed mock catalogues of the Dark Energy Survey. We implement: 1) second and third moments; 2) wavelet phase harmonics (WPH); 3) the scattering transform (ST). Our analysis is fully based on simulations, it spans a space of seven $\nu w$CDM cosmological parameters, and it forward models the most relevant sources of systematics of the data (masks, noise variations, clustering of the sources, intrinsic alignments, and shear and redshift calibration). We implement a neural network compression of the summary statistics, and we estimate the parameter posteriors using a likelihood-free-inference approach. We validate the pipeline extensively, and we find that WPH exhibits the strongest performance when combined with second moments, followed by ST, and then by third moments. The combination of all the different statistics further enhances constraints with respect to second moments, up to 25 per cent, 15 per cent, and 90 per cent for $S_8$, $\Omega_{\rm m}$, and the Figure-Of-Merit ${\rm FoM_{S_8,\Omega_{\rm m}}}$, respectively. We further find that non-Gaussian statistics improve constraints on $w$ and on the amplitude of intrinsic alignment with respect to second moments constraints. The methodological advances presented here are suitable for application to Stage IV surveys from Euclid, Rubin-LSST, and Roman with additional validation on mock catalogues for each survey. In a companion paper we present an application to DES Year 3 data.
\end{abstract}

\author{M.~Gatti}\email{marcogatti29@gmail.com}
\affiliation{Department of Physics and Astronomy, University of Pennsylvania, Philadelphia, PA 19104, USA}
\author{N. Jeffrey}
\affiliation{Department of Physics \& Astronomy, University College London, Gower Street, London, WC1E 6BT, UK}
\author{L. Whiteway}
\affiliation{Department of Physics \& Astronomy, University College London, Gower Street, London, WC1E 6BT, UK}
\author{J. Williamson}
\affiliation{Department of Physics \& Astronomy, University College London, Gower Street, London, WC1E 6BT, UK}
\author{B. Jain}
\affiliation{Department of Physics and Astronomy, University of Pennsylvania, Philadelphia, PA 19104, USA}
\author{V. Ajani}
\affiliation{Department of Physics, ETH Zurich, Wolfgang-Pauli-Strasse 16, CH-8093 Zurich, Switzerland}
\author{D. Anbajagane}
\affiliation{Kavli Institute for Cosmological Physics, University of Chicago, Chicago, IL 60637, USA}
\affiliation{Department of Astronomy and Astrophysics, University of Chicago, Chicago, IL 60637, USA}
\author{G. Giannini}
\affiliation{Kavli Institute for Cosmological Physics, University of Chicago, Chicago, IL 60637, USA}
\author{C. Zhou}
\affiliation{ Santa Cruz Institute for Particle Physics, Santa Cruz, CA 95064, USA}
\author{A. Porredon}
\affiliation{Ruhr University Bochum, Faculty of Physics and Astronomy, Astronomical Institute, German Centre for Cosmological Lensing, 44780 }
% ...
\collaboration{DES Collaboration}

\vspace{0.2cm}


\preprint{DES-2023-0634}
\preprint{FERMILAB-PUB-23-634-PPD}


\maketitle

\setcounter{footnote}{1}


\section{Introduction}

Weak gravitational lensing is a powerful tool for studying the large-scale structure  (LSS) of the mass distribution in the Universe. Photons emitted by distant galaxies are deflected when passing through regions of spacetime affected by the mass distribution between the sources and the observer \citep{Einstein1936}. By measuring the shapes of numerous galaxies, statistical methods enable us to deduce the projected spatial distribution of the mass responsible for these weak deflections and thereby create weak lensing \textit{mass maps} (\citealt{VanWaerbeke2013, Vikram2015, Chang2015, Liu2015, Chang2018, Oguri2018}; \citealt*{y3-massmapping}). At the time of writing, ongoing and upcoming surveys, including the Dark Energy Survey (DES, \citealt{DES2016}), the Kilo-Degree Survey (KIDS, \citealt{Kuijken2015}), the Hyper Suprime-Cam (HSC, \citealt{Aihara2018}), the Vera C. Rubin Observatory's Legacy Survey \citep{Abell2009}, and the Euclid mission \citep{Laureijs2011}, are measuring (or being readied to measure) galaxy shapes on a massive scale, encompassing thousands of square degrees across the sky. Notably, the DES project recently measured the shapes of more than 100 million galaxies in an area of approximately 5000 square degrees in the southern hemisphere \citep*{y3-shapecatalog}, which led to the production of the most extensive weak lensing mass map from a galaxy survey to date \citep*{y3-massmapping}. In parallel, measurements of the lensing of the cosmic microwave background (CMB) have led to maps of the mass distribution projected all the way to the redshift of the last scattering surface (e.g. \citealt{Madhavacheril2023}).

If a mean-zero random field is Gaussian, then a two-point statistic captures all its statistical information. Two-point statistics of the shear field can be measured in harmonic, configuration, or other spaces:  e.g. power spectra (harmonic space), shear two-point correlation function (configuration space), or COSEBI (Complete Orthogonal Sets of $E/B$-Integrals) have to date been measured and used for cosmological parameter estimation (e.g. \citealt{Asgari2021}; \citealt{y3-cosmicshear1}; \citealt*{y3-cosmicshear2}; \citealt{Doux2022,Dalal2023,Li2023}). 
However, a significant amount of the information contained in weak lensing mass maps lies in their non-Gaussian features, and these features are not fully captured by two-point statistics. Many recent studies, using a wide range of tools and statistics, have tried to extract the non-Gaussian information; examples include higher-order moments \citep{VanWaerbeke2013,Petri2015,Vicinanza2016,Chang2018,Vicinanza2018,Peel2018, G20,moments2021,Porth2021},  peak counts \citep{Dietrich2010, Kratochvil2010, Liu2015, Kacprzak2016, Martinet2018, Peel2018, Shan2018, ajani_peaks,Zuercher2021,HD2022,Zuercher2022}, one-point probability distributions \citep{Barthelemy2020,Boyle2021,Thiele2020}, Minkowski functionals \citep{Kratochvil2012,Petri2015,Vicinanza2019,Parroni2020,Grewal2022},  Betti numbers \citep{Feldbrugge2019,Parroni2021}, persistent homology \citep{Heydenreich2021,Heydenreich2022}, scattering transform coefficients \citep{Cheng2020,Valogiannis2022a,Valogiannis2022}, wavelet phase harmonic moments \citep{Allys2020}, kNN and CDFs \citep{Anbajagane2023,Banerjee2023},  map-level
inference \citep{Porqueres2022,Boruah2022}, and machine-learning methods \citep{Ribli2018, Fluri2018,Fluri2019,jeffrey_lfi,Lu2023}. Many of these studies, however, are limited to being proofs of concept, restricted to idealized simulated scenarios (due to the challenges associated with applying these techniques to real-world data). Nevertheless, the field is rapidly progressing, with a number of recent applications to observational data \citep{Liu2015, Kacprzak2016,Martinet2018,Fluri2019,jeffrey_lfi,moments2021,Zuercher2022,Heydenreich2022,Fluri22,Lu2023}.

One of the major challenges in exploiting non-Gaussian statistics is the need for accurate  modelling of measurements. Analytic models are available only for a small set of non-Gaussian summary statistics (e.g. moments), and often these models are reliable only at large scales. Consequently, many studies resort to using simulations to forward model the observables. This procedure introduces its own challenges. Most importantly, computational resources are a significant concern, as it is necessary to run numerous $N$-body simulations to cover the parameter space explored in the analysis. Additionally, it is a formidable task to incorporate all the relevant observational and systematic effects into these simulations. Finally, it is critical to estimate efficiently the parameter posteriors; this requires techniques able to recover accurately the posterior from a limited number of simulation samples (specifically, those available at the locations in parameter space of the $N$-body simulations).


In this study, we use a set of non-Gaussian summary statistics of weak lensing mass maps to constrain cosmology with the first three years (Y3) of  data from DES. This work validates the methodology using simulations; a companion paper applying this framework to the DES Y3 data will follow. Our analysis makes use of the following Gaussian and non-Gaussian statistics: second and third-order moments, wavelet phase harmonic (WPH) moments, and the scattering transform (ST) coefficients. Moments have previously been used in analysing DES data using analytical models instead of simulations \citep{moments2021}; in contrast, this paper fully relies on a simulation based inference. {Furthermore, WPH moments and the ST have not been applied to data before. The WPH moments are second moments of smoothed weak lensing mass maps that have undergone a non-linear transformation, allowing for the exploration of the non-Gaussian features of the field. The ST coefficients are built through a series of smoothing and modulus operations applied to the input field, followed by an average. WPH and ST have two advantages relative to traditional higher-order correlations: better constraining power and (as they do not go to higher than second order in the field) lower sensitivity to noise fluctuations \citep{Allys2020}. WPH and ST are frequently compared to convolutional neural networks (CNNs) because their definition bears similarities to the architecture of CNNs \citep{Mallat_2016}; however, their definition depends only on a handful of parameters (parameters that have clear physical interpretation), and, in contrast to CNNs, they require no training.}

For this work we produced a set of $N$-body simulations (Jeffrey et al., in prep.) that explores a seven-dimensional parameter space. The simulations incorporate key observational and astrophysical systematic effects impacting weak lensing analyses, including photometric redshift uncertainties, shear calibration errors, intrinsic alignments, and source clustering (as described in \cite{Gatti2023}, this latter effect has a greater influence on non-Gaussian statistics than on Gaussian statistics). To obtain posterior estimates of the parameters, we employ an optimal data compression technique called neural compression, which significantly reduces the dimensionality of our summary statistics. Subsequently, we employ a likelihood-free inference (LFI, e.g. \citealt{jeffrey_lfi}) approach, enabling us to estimate posteriors without imposing restrictive assumptions about the likelihood or data model. This powerful approach circumvents various technical challenges associated with conventional analysis methods, such as covariance matrix estimation and sampling from high-dimensional Bayesian hierarchical models. We also examine the combination of the three non-Gaussian summary statistics considered in this work; to date, the combination of distinct non-Gaussian summary statistics has only been explored in idealized simulations \citep{Zuercher2022,euclid2023}, and its application to real data remains unexplored. We test the methodology extensively with simulated data to ensure that the results from survey data are unbiased. 

This paper is organised as follows. Section 2 summarizes the survey data as well as the simulations used for our model predictions and for validation. Section 3 describes the various summary statistics, their covariances, and the compressed statistics obtained from them. We describe and validate in Section 4 the LFI methodology for parameter inference and in Section 5 the choice of scale cuts. Section 6 validates the full pipeline with an end-to-end simulated cosmological analysis, and we summarise our results in Section 7. 

\section{Data and simulations}

\subsection{DES Y3 weak lensing catalogue}

We use the DES Y3 weak lensing catalogue \citep*{y3-shapecatalog}; this contains 100,204,026 galaxies, with a weighted $n_{\rm eff}=5.59$~galaxies~arcmin$^{-2}$, over an effective area of 4139  deg$^2$. It was created using the \textsc{METACALIBRATION} algorithm \citep{HuffMcal2017, SheldonMcal2017}, which provides self-calibrated shear estimates starting from (multi-band) noisy images of the detected objects. A residual small calibration (in the form of a multiplicative shear bias) is provided; it is based on sophisticated image simulations \citep{y3-imagesims} and it accounts for blending-related detection effects. An inverse variance weight is further assigned to each galaxy in the catalogue to enhance the overall signal-to-noise. The sample is divided into four tomographic bins of roughly equal number density \citep*{y3-sompz} and redshift distributions are provided by the SOMPZ method  \citep*{y3-sompz} in combination with clustering redshift constraints \citep*{y3-sourcewz} and corrections due to the redshift-dependent effects of blending \citep{y3-imagesims}.

\subsection{Simulations}

\begin{table}
\caption {Distribution of the parameters spanned by the Gower St mock catalogues (second column), and the prior used in the cosmological analysis (third column). The prior used in the analysis can differ from the distribution of the samples as long as these parameters have been explicitly used during the training of the Neural Density Estimators (NDEs) when learning the likelihood surface; more details are given in \S \ref{sect:LFI}. For the third column, we report the analysis prior only if it is different from the mocks parameters distribution.  }
\centering
\begin{tabular}{c|c|c}
\hline
\textbf{Parameter} & \textbf{Mocks parameters} & \textbf{Analysis prior} \\
 & \textbf{distribution} & \\
\hline 
\hline 
$\Omega_{\rm m}$ & mixed active-learning  & $ \mathcal{U}(0.15,0.52)$ \\ & in $\mathcal{U}(0.15,0.52)$   \\ 
\hline
$S_8$ & mixed active-learning  & $\mathcal{U}(0.5,1.0)$ \\& in $\mathcal{U}(0.5,1.0)$ \\
\hline
\hline
$w$ & $\mathcal{N}(-1,\frac{1}{3})$ for $-1<w<-\frac{1}{3}$ & $\mathcal{U}(-1,\frac{1}{3})$ \\
& $0$ else  \\
\hline
$n_s$ & $\mathcal{N}(0.9649, 0.0063)$   \\
\hline
$h$ & $\mathcal{N}(0.7022, 0.0245)$ \\
\hline
$\Omega_{\rm b}h^2 $&  $ N(0.02237, 0.00015)$\\
\hline
$\log(m_{\nu}) $& $ \mathcal{U}[\log(0.06), \log(0.14)]$ \\
\hline
\hline
$A_{IA}$ & $\mathcal{U}[-3, 3]$  \\
\hline
$\eta_{IA}$ & $\mathcal{U}[-5, 5]$ \\
\hline
$m_{1}$ & $\mathcal{N}(-0.0063,0.0091)$ \\
\hline
$m_{2}$ & $\mathcal{N}( -0.0198,0.0078)$ \\
\hline
$m_{3}$ & $\mathcal{N}( -0.0241,0.0076)$ \\
\hline
$m_{4}$ & $\mathcal{N}(-0.0369, 0.0076)$ \\
\hline
$\bar{n}_i(z)$ & $p_{\textsc{HyperRank}}(\bar{n}_i(z) | x_{\rm phot})$  \\ 
\hline
\end{tabular}
\label{parameter}
\end{table}

\begin{figure}
\includegraphics[width=0.45 \textwidth]{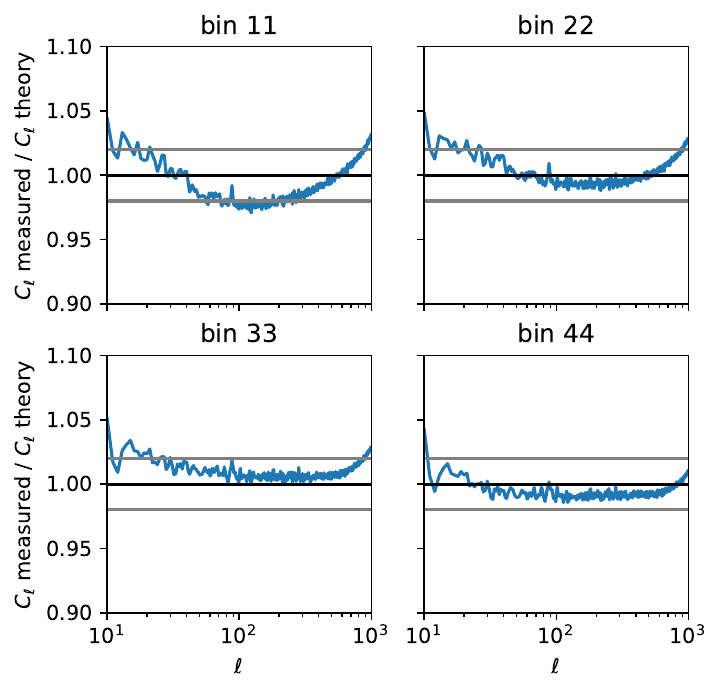}
\caption{Ratio of the convergence power spectrum ($C_{\ell}$) as measured in the Gower St simulations and that from theoretical predictions. The power spectrum has been measured on full-sky, noiseless convergence maps. The ratio has been averaged over all the simulations available. The two horizontal lines are provided for reference and show that the typical deviation is at the 2 percent level.}
\label{fig:Dirac_tests}
\end{figure}

\begin{figure*}
\includegraphics[width=0.95 \textwidth]{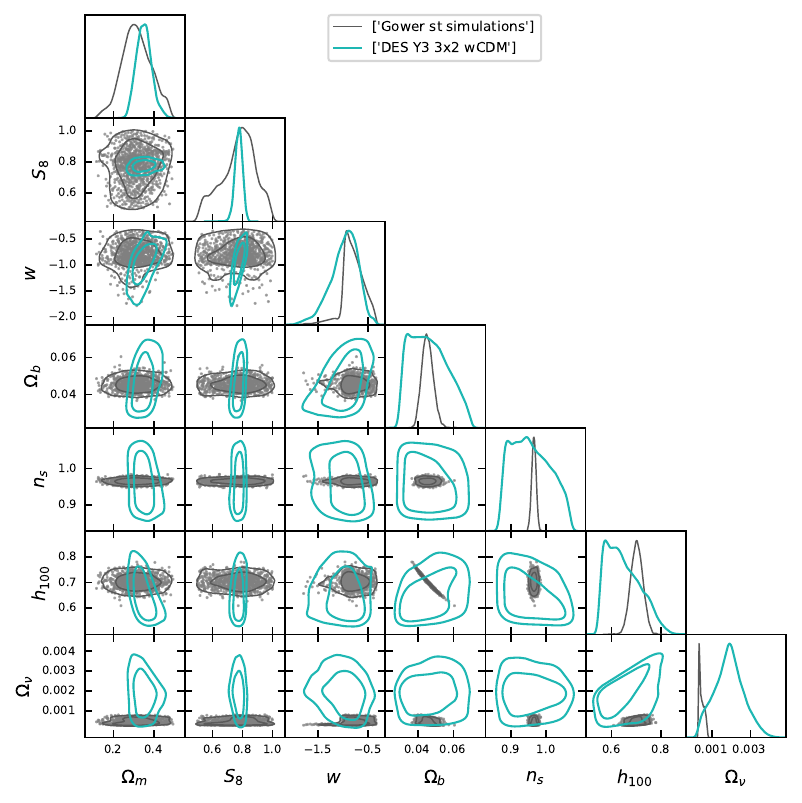}
\caption{Distribution of Gower St simulations for the seven parameters spanned in this analysis (grey). The two-dimensional marginalised contours in these figures show the 68 per cent and 95 per cent percentile of the simulations. For comparison purposes, we also show (cyan) the posterior from the DES Y3 3x2 $\nu w$CDM analysis \citep{y3-3x2ptkp}.}
\label{fig:Dirac_sims}
\end{figure*}

\subsubsection{Gower St simulations}\label{sect:Dirac}
We use the Gower St simulation suite (Jeffrey et al., in prep.) to build our pipeline. The suite consists of 791 gravity-only full-sky $N$-body simulations, produced using the \textsc{PKDGRAV3} code \citep{potter2017pkdgrav3}. The simulations span a seven-dimensional parameter space in $\nu w$CDM ($\Omega_{\rm m}$, $\sigma_8$, $n_s$, $h_{100}$, $\Omega_{\rm b}$, $w$, $m_{\nu}$). The parameter space is not spanned uniformly (Fig.~\ref{fig:Dirac_sims} shows the simulation distribution in the parameter space). $\Omega_{\rm m}$ and $\sigma_{\rm 8}$ have been sampled with a mixed active-learning strategy; in particular they were at first distributed according to the existing DES analysis constraints, and then, after an initial simple blind power spectrum analysis, new simulations were run with $\sigma_8$ and $\Omega_{\rm m}$ values (known only to the computer) in regions of parameter space with poor accuracy of the likelihood estimates (see Jeffrey et al., in prep.). The other parameters were chosen to be distributed as follows:
\begin{itemize}
\item{$n_s \sim \mathcal{N}(0.9649, 0.0063)$; from Planck \citep{aghanim2020planck} but with the standard deviation boosted by a factor of 1.5.} 
\item{$h \sim \mathcal{N}(0.7022, 0.0245)$; consistent with both SH0ES~\citep{Riess_2022} and Planck \citep{aghanim2020planck}.}
\item{$\Omega_{\rm b}h^2 \sim N(0.02237, 0.00015)$; from Planck \citep{aghanim2020planck}.}
\item{$w \sim \mathcal{N}(-1, 1/3)$, but with values less than $-1$ or greater than $-1/3$ then discarded. For a few (64) simulations, part of the `science verification' runs, this discarding was not done. We kept these simulations during the training of our NDEs, but we used a hard prior at $w>-1$ for the analysis.}
\item{$m_{\nu}$: fixed at $0.06$ for 192 simulations and with $\log(m_{\nu}) \sim \mathcal{U}[\log(0.06), \log(0.14)]$ thereafter.}
\end{itemize}
In the above, $\mathcal{N}(\mu,\sigma)$ denotes a normal distribution with the indicated mean and standard deviation and $\mathcal{U}[a, b]$ denotes a uniform distribution with the indicated limits. {We note that the sampling strategy does not necessarily affect our posteriors; more details are given in \S~\ref{sect:LFI}}.

The simulations used up to ten replicated boxes in each direction so as to span the redshift interval from $z = 0$ to $z = 49$, although note that the bulk of our redshift distributions ($z<1.5$) can be covered by only three replications. Each individual box contains $1080^3$ particles and has a side-length of 1250 $h^{ - 1}$ Mpc. For each simulation, lens planes $\delta_{\rm shell}(\hat{\boldsymbol{\rm n}}, \chi)$ are provided at $\sim 100$ redshifts from $z=49$ to $z=0.0$, equally spaced in proper time. For this work, we downsample the original resolution of \textsc{NSIDE} = 2048 to \textsc{NSIDE} = 512 (with pixel size $\approx$ 6.9 arcmin).  The lens planes are provided as \healpix{} \citep{GORSKI2005} maps and are obtained from the raw number particle counts.\footnote{$\delta_{\rm shell}(\hat{\boldsymbol{\rm n}}, \chi) = n_{\rm p}(\hat{\boldsymbol{\rm n}}, \chi)/ \langle n_{\rm p}(\hat{\boldsymbol{\rm n}}, \chi) \rangle-1$, where $\langle \rangle$ indicates the spatial average and $n_{\rm p}$ is the number of particles in a given pixel $p$.} The lens planes are converted into convergence planes $\kappa_{\rm shell}(\hat{\boldsymbol{\rm n}}, \chi)$ under the Born approximation (e.g. Eq.~2 from \citealt{Fosalba2015}). Lastly, shear planes $\gamma_{\rm shell}(\hat{\boldsymbol{\rm n}}, \chi)$ are obtained from the convergence maps using a full-sky generalisation of the \cite{KaiserSquires} algorithm \citep*{y3-massmapping}.

{We validate the Gower St simulations by comparing the power spectra measured on the full-sky convergence maps, weighted by the DES redshift distributions, against theory predictions obtained using \texttt{halofit} \citep{Takahashi2012}. Note that we did not use the more recent (and more accurate) \texttt{EuclidEmu} \citep{Euclidemu2021} for this comparison, as \texttt{EuclidEmu} covers only a very limited portion of our parameter space. We generally do not expect an agreement better than 2 per cent, as this is the typical relative error between different non-linear power spectrum prescriptions or other modelling implementations (e.g. neutrinos). At the largest scales, on the other hand, box-size effects and/or cosmic variance in the simulations might impact the comparison}. To perform the test, we build the redshift weighted convergence maps as
\begin{equation}
\kappa(p) = \frac{\sum_s \bar{n}(s)  \kappa(p, s)}{\sum_s \bar{n}(s)}, 
\end{equation}
where $p$ is a map pixel, $s$ is the redshift shell, $\kappa(p, s)$ is the noiseless convergence field from the simulation, and $\bar{n}(s)$ is the DES galaxy count across the whole footprint \citep*{y3-sompz}. For each of the four DES tomographic bins, we computed the ratio between the power spectrum of the simulated convergence field $\kappa(p)$ and the theory predictions from \texttt{halofit}. We show the average of the ratio over all the Gower St simulations in Fig.~\ref{fig:Dirac_tests}; the agreement is good, within 2 percent over the range of multipoles considered in this work (up to $\ell=1024$; see \S \ref{sect:sumst}).

\subsubsection{CosmoGridV1 simulations}
We use a subset of the simulations from the \texttt{CosmoGridV1} suite \citep{cosmogrid1} for additional testing and to determine the scale cuts that need to be removed because of baryonic contamination. The \texttt{CosmoGridV1} simulations have been produced using the \textsc{PKDGRAV3} code \citep{potter2017pkdgrav3}. From the available \texttt{CosmoGridV1} simulations we chose a set of one hundred full-sky simulations at the fiducial cosmology $\sigma_8 = 0.84$, $\Omega_{\rm m}=0.26$, $w=-1$, $H_0=67.36$, $\Omega_{\rm b}=0.0493$, $n_{\rm s}=0.9649$. Each individual simulation has also been post-processed with a baryonification algorithm that mimics the impact of baryons at small scales. 
The algorithm used is the baryonic correction model \citep{Schneider2015,Arico2020}, which adjusts the particle positions in gravity-only simulations to mimic the impact of various baryonic processes on the density distribution. The cosmology has been chosen to be centred well within our priors for $\sigma_8$, $\Omega_{\rm m}$ and $w$. {The baryonic correction model depends on several parameters (up to seven); these impact both the shape and the amplitude of the power spectrum. The parameter that has the largest impact is $M_c$, the mass scale at which haloes have lost half of their gas. A value of $M_c = 10^{13.82} M_{\odot}$ has been adopted, following \cite{Fluri22,Schneider2019}, based on observed X-ray gas fractions. The values of the other parameters have been estimated  by comparing against current X-ray observation; see \cite{Schneider2019} model \texttt{B-avrg} for a list of the values. More details are given in \S \ref{sect:scale_cuts}, where we evaluate the impact of baryons on our constraints.} 

The simulations were obtained using multiple replicated boxes in each direction so as to span the redshift interval from z = 0 to z = 3.5. Each individual box contains $832^3$ particles and has a side-length of 900 $h^{ - 1}$ Mpc.  For each simulation, lens planes $\delta_{\rm shell}(\hat{\boldsymbol{\rm n}}, \chi)$ are provided at $\sim 69$ redshifts from $z=3.5$ to $z=0.0$, equally spaced in proper time.  We downsample the original resolution of \textsc{NSIDE} = 2048 to \textsc{NSIDE} = 512 (with pixel size $\approx$ 6.9 arcmin).  Last, convergence and shear planes are obtained using the same procedure as adopted for the Gower St simulations.

\subsubsection{DES Y3 maps-making procedure}\label{sect:map_making_procedure}

We use the simulated full-sky convergence maps to generate DES Y3-like weak lensing convergence maps following the procedure outlined in \cite{Gatti2023}. The procedure is similar to others used in past DES analyses (e.g. \citealt{moments2021,Zuercher2021b}), but improves upon them by introducing for the first time an efficient recipe to forward model source clustering effects. We further extend that procedure to incorporate extra observational systematic effects. Let $p$ be a pixel, $s$ a thin redshift shell, $\gamma(p, s)$ the noiseless shear from the shear simulation, and $\bar{n}(s)$ the galaxy count across the whole footprint \citep*{y3-sompz}. Additionally, let $m$ be the multiplicative shear bias that models shear measurement uncertainties \citep{y3-imagesims}, and let $\gamma_{\rm IA}(p, s)$ be the intrinsic alignment contribution to each pixel.  Let $\delta(p, s)$ be the matter overdensity in the shear simulation, and let $b_g$ be the galaxy-matter bias of the weak lensing sample. Each galaxy has a shear weight $w_g$ and ellipticity $e_g$. We randomly rotate the DES galaxy ellipticities to erase the cosmological signal of the catalogue.

The mock shear signal in pixel $p$ is set to
\begin{multline}
\label{eq:sc_pixel}
\gamma(p) = \frac{\sum_s \bar{n}(s) [1 + b_g \delta(p, s)] (1 + m) [\gamma(p, s)+\gamma_{\rm IA}(p, s)]}{\sum_s \bar{n}(s) [1 + b_g \delta(p, s)]}  + \\
\left(\frac{\sum_s \bar{n}(s)}{\sum_s \bar{n}(s) \left[1 + b_g \delta(p, s)\right]}\right)^{1/2} F(p) \, \frac{\sum_g w_g e_g}{\sum_g w_g}.
\end{multline}

The signal term is a weighted average over shells; here the weights have been amended to include a shear-correlated source galaxy count \citep{Gatti2023}.  The term $F(p)$ in Eq.~\ref{eq:sc_pixel} is a near-unity scale factor introduced to avoid double-counting source clustering effects, adjusting the even moments of the noise of the maps, as the DES Y3 catalogue used to model the shape noise of the pixels is already affected by source clustering. We follow \cite{Gatti2023} and assume
\begin{equation}
    F(p) = A\sqrt{1-B \sigma_{e}^2(p)},
\end{equation}
where $\sigma_{e}^2(p)$ is the variance of the pixel noise and $A = [0.97,0.985,0.990,0.995]$ and $B = [0.1,0.05,0.035,0.035]$ are constants (one for each tomographic bin). A further validation of the noise properties of our simulations is provided in Appendix \ref{sect:noise_terms}. The intrinsic alignment term $\gamma_{\rm IA}(p, s)$ is
\begin{equation}
    \label{eq:IA_}
    \gamma_{\rm IA}(p, s) = A_{\rm IA} \left(\frac{1+z}{1+z_0} \right)^{\eta_{\rm IA}} \frac{c_1 \rho_{\rm crit} \Omega_{m}}{D(z)} S(p,s),
\end{equation}
with $z_0= 0.62$, $c_1=5\times 10^{-14}M_{\odot}h^{-2}$Mpc$^2$ (\citealt{Bridle2007}), $\rho_{\rm crit}$ the critical density, $D(z)$ the linear growth factor, and $S(p,s)$ the shear tidal field. We obtain $S(p,s)$ directly from the density field $\delta(p, s)$ by applying the (inverse) Kaiser-Squires algorithm. The two intrinsic alignment parameters $A_{\rm IA}$ and $\eta_{\rm IA}$ in Eq.~\ref{eq:IA_} control respectively the amplitude and the redshift evolution of the intrinsic alignment signal.
In writing Eq.~\ref{eq:IA_} we have followed the non-linear alignment model (NLA, \citealt{Bridle2007}); however, since we are including source clustering in our simulations (the $(1+b_{g})\delta(p,s)$ term in Eq.~\ref{eq:sc_pixel}), the final intrinsic alignment model includes extra clustering terms beyond the original NLA implementation. These terms are similar to the clustering term included in the tidal-torque alignment (TATT) model \citep{blazek19}; that paper, however, estimates those contributions only for catalogue-based Gaussian statistics using tree-level perturbation theory, whereas our implementation directly uses the clustering of the simulation and generalises to all the summary statistics considered in this work. {With the simulations at hand, we were not able to include a more sophisticated IA model (e.g, including all the terms of the TATT model, as was done for the fiducial DES Y3 weak lensing analysis of \citealt{y3-cosmicshear1}, \citealt*{y3-cosmicshear2}). However, we note that the DES Y3 cosmological analyses on data (\citealt{y3-cosmicshear1}, \cite*{y3-cosmicshear2}, \cite{y3-3x2ptkp}) have not yielded any substantial indications favouring the adoption of a more complex model (such as TATT) over NLA; moreover, these results are consistent with a zero intrinsic alignment amplitude. For these reasons, we consider the IA model implemented here to be adequate for our analysis.}

\begin{figure}
\includegraphics[width=0.45\textwidth]{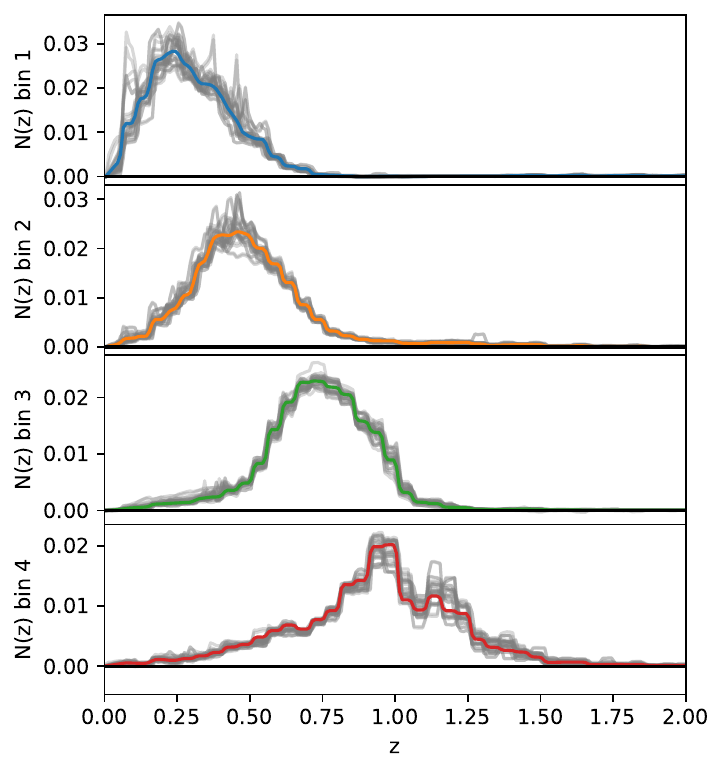}
\caption{Redshift distributions as estimated in data for the four DES Y3 tomographic bins \citep*{y3-sompz}. The solid coloured lines are the average $n(z)$ for each bin; the grey lines are a few ($\sim 50$) samples encompassing the redshift calibration uncertainties and that we use to create the mocks.}
\label{fig:nz_comp}
\end{figure}

{This procedure is repeated for each of the four tomographic bins of the DES Y3 source catalogue. As we can cut four independent DES Y3 footprints from each full-sky map, we produce a total of 3164 independent DES Y3 shear mock maps from the Gower St simulations. Additionally, we produced another $\sim$9492 pseudo-independent DES Y3 shear mock maps by shifting the four independent DES Y3 footprints by 45, 90, and 135 degrees, so as to cover slightly different parts of the full-sky maps we generated, for a total of 12656 mocks. We used these mocks to train the neural network compression of the summary statistics. Then we repeated this whole procedure to generate another 12656 mocks, with different shape noise, that we used to train the neural density estimators used for the likelihood-free inference. In total, therefore, we produced 25312 pseudo-independent mocks.}

The process of creating mock datasets involves a number of unconstrained parameters, including four multiplicative shear biases, four redshift distributions, and the intrinsic alignment parameters. When generating each of the 25312 pseudo-independent mocks, we select one of these parameters randomly from their respective priors (as detailed in Table~\ref{parameter}). For the redshift distributions, for each mock we pick at random one of the multiple realisations provided by the \textsc{hyperrank} methodology \citep{y3-hyperrank} using photometric redshift data $x_{\rm phot}$; we then use it as a $\bar{n}(s)$. These realisations encompass the redshift calibration uncertainties. In Fig.~\ref{fig:nz_comp}, we present for each tomographic bin a few of the realisations used in this study. Finally, we used the 100 independent \texttt{CosmoGridV1} full-sky realisations to generate two sets (with and without baryonic feedback effects) of 400 independent DES Y3 shear mock maps.

\section{Summary statistics}\label{sect:sumst}

We use different Gaussian and non-Gaussian weak lensing summary statistics in this work. All the summary statistics are applied to weak lensing mass maps; as a first step, therefore, we create the weak lensing mass maps starting from the shear maps. This is achieved by using a full-sky generalisation of the \cite{KaiserSquires} algorithm \citep*{y3-massmapping}. This produces noisy weak lensing mass maps in the form of \healpix{} maps with a resolution of \textsc{NSIDE} = 512 (corresponding to a pixel size of $\approx$ 6.9 arcminutes). {This procedure is repeated for all four tomographic bins of our catalogue. During the creation of the mass maps, we further applied a cut at $\ell_{\rm max} = 1024$. The maps at \textsc{NSIDE} = 512 formally have non-zero support up to $\ell=1535$; most of their power, however, is suppressed above $\ell ~\sim 1000$ because of the pixel window function. We chose to incorporate this particular cut when we were constructing the pipeline as we were assuming then that we would need to remove these scales due to potential baryonic contamination; we did not revisit this choice after the scale cut test presented in \S \ref{sect:scale_cuts}, as it would have required us to redo the creation of the mocks and measurements.}

The summary statistics considered in this work are: 1) second and third moments; 2) wavelet phase harmonics; 3) the scattering transform. The statistics are applied to smoothed versions of the weak lensing maps, with the type of smoothing depending on the statistic: moments use top hat filters, while wavelet phase harmonics and the scattering transform use wavelet filters \citep{Cohen1995,Mallat1999,VDB1999}. In all cases we smooth the maps using filters with different sizes. More details and relevant equations are presented below. 

\subsection{Second and Third moments}\label{sect:2nd3rd}

The first statistics considered are second and third moments of the weak lensing mass maps \citep{VanWaerbeke2013,Petri2015,Vicinanza2016,Chang2018,Vicinanza2018,Peel2018, G20,moments2021}. While second moments are a Gaussian statistic, third moments probe additional non-Gaussian information of the field. Second and third moments of the DES Y3 weak lensing mass maps have been recently used in \cite{moments2021} to infer cosmology; here we adopt that paper's implementation of the moments estimator.

We first smooth the maps using a top-hat filter with different smoothing scales. {In practice, this is achieved by multiplying the coefficients of the harmonic decompositions of the weak lensing mass maps by}
\begin{equation}
\label{eq:filter}
W_{\ell}(\theta_0) = \frac{P_{\ell-1}({\rm cos}(\theta_0))-P_{\ell+1}({\rm cos}(\theta_0))}{(2\ell+1)(1-{\rm cos}(\theta_0))},
\end{equation}
where $P_{\ell}$ is the Legendre polynomial of order $\ell$, $\theta_0$ is the smoothing scale, and $\ell$ is the multipole. We consider eight smoothing scales equally (logarithmically) spaced from $8.2$ to $221$ arcmin, and we denote the smoothed lensing mass map of tomographic bin $i$ by $\kappa_{\theta_0,p}^i$. We estimate the second and third moments as follows:
\begin{equation}
\avg{\hat{\kappa}^2_{\theta_0}}(i, j) = \Avg_p \left( \kappa_{\theta_0,p}^i \, \kappa_{\theta_0,p}^j \right)
\end{equation}
\begin{equation}
\avg{\hat{\kappa}^3_{\theta_0}}(i, j, k) = \Avg_p \left( \kappa_{\theta_0,p}^i \, \kappa_{\theta_0,p}^j \, \kappa_{\theta_0, p}^k \right) .
\end{equation}
Here $i,j,k$ refer to different tomographic bins; all combinations of tomographic bins are considered (ten independent combinations for second moments and 20 for third moments). The average is over all pixels $p$ on the full sky (i.e. $1 \le p \le N_{\textrm{tot}}$), including regions outside the footprint, since the Kaiser-Squires conversion, and the subsequent smoothing, transfers some of the signal from inside to outside the DES footprint. 

We can only estimate \textit{noisy} realisations of the weak lensing mass maps: $\kappa_{{\rm obs}} = \kappa + \kappa_{{\rm N}}$. Any statistic measured with data will include noise contributions  \citep{VanWaerbeke2013}. {When comparing measurements to analytical predictions, noise-only terms are normally subtracted to ease the comparison. If the noise-only terms are estimated from the data (via, for example, random rotation of the ellipticity measurements), subtracting the noise terms can increase the measurement uncertainties, because the noise terms estimates are affected by shot noise. While it would be possible to have multiple estimates of the noise terms for every map to reduce the shot noise contribution, we simply chose to not subtract these terms, except in a few particular cases.}

As for moments, we decided only to subtract the following specific noise terms from our third moments estimator:

\begin{equation}
\avg{\hat{\kappa}^3_{\theta_0}} = \avg{\hat{\kappa}^3_{\theta_0, \rm obs}}-\avg{\hat{\kappa}_{\theta_0,{ \rm obs}}\hat{\kappa}^2_{\theta_0, \rm N}};
\end{equation}
%
For third moments, we subtracted noise-signal third moments of the form $\avg{\hat{\kappa}_{\theta_0, \rm obs}\hat{\kappa}^2_{\theta_0, \rm N}}$. These terms are strictly non-zero because of spurious noise-signal correlations arising from source clustering; \citep{Gatti2023} found that subtracting these terms reduces the impact of source clustering (and hence potential biases in the analysis if the source clustering is mismodelled in simulations). Other terms ($\avg{\hat{\kappa^2}_{\theta_0, \rm obs}\hat{\kappa}_{\theta_0, \rm N}}$ and $\avg{\hat{\kappa}^3_{\theta_0, \rm N}}$) were not subtracted as they average to zero even in presence of source clustering \citep{Gatti2023}.

\subsection{Wavelet Phase Harmonics}\label{sect:WPH} Wavelet phase harmonics \citep{Mallat_2016,Allys2020} are the second moments of smoothed weak lensing mass maps that have undergone a non-linear transformation. The fields are first smoothed using a directional, multi-scale wavelet transform \citep{Cohen1995,Mallat1999,VDB1999}; the wavelets have the advantage of being localised both in Fourier and real space, contrary to the top-hat filters used in this work for the second and third moments, which are local only in real space. Moreover, we adopt `directional' wavelets, instead of using an isotropic filter.

We use the package \textsc{PYWPH} \footnote{\url{https://github.com/bregaldo/pywph}} to smooth our maps. The package works on a two-dimensional projection rather than on a sphere. Therefore, we first cut multiple square patches of roughly 14.6 degrees of side covering the full DES footprint. For this we use a gnomonic projection (as implemented in the \healpix{} \textsc{gnomview} function), converting our patches to a 128x128 pixelated grid with a pixel scale of 6.8 arcminutes. Due to projection effects, the same portion of a map might appear in multiple projected patches; we mask pixels accordingly to avoid double-counting. Note that both simulated and real data maps undergo the same projection process.

We then smooth the projected patches using `bump steerable wavelets'. Begin in Fourier space, where we define the wavelet
\begin{multline}
\hat{\psi} (\vec{k}) = 
\begin{cases}
    0.7309 \,{\rm exp}\left( \frac{-(|\vec{k}|-\xi_0)^2}{\xi_0^2-(|\vec{k}|-\xi_0)^2}\right) { \cos^2(\arg(\vec{k}))} \vspace{5pt} \\  \qquad \textrm{ if } 0<|\vec{k}|< 2 \xi_0 \textrm{ and } k_x \ge 0 \vspace{5pt},\\
    0 \quad \textrm{otherwise}.
\end{cases}
\end{multline}
Here $\vec{k} = (k_x, k_y)$ is the two-dimensional Fourier wavenumber, while $\xi_0$ denotes the central frequency of the wavelet (the full vector is $\vec{\xi_0} = (\xi_0,0)$) and is set to $\xi_0=0.85 \pi$ following \cite{Mallat2020}; the prefactor and the power of the cosine function corresponds to $L=3$ in their notation. Note that $\hat{\psi}$ has finite width (i.e. `compact support') in Fourier space. The real space Fourier transform $\psi$ of this is then our `mother' wavelet, from which other wavelets can be obtained by dilating and rotating:
\begin{equation}
    {\psi}_{n,\ell}\left(\vec{\theta} \right)= 2^{-2n}{\psi}\left( 2^{-n}\Rot_{-\ell} \vec{\theta}\right).
\end{equation}
Here $\Rot_{\ell}$ denotes rotation by an angle $\pi \ell / L$; we consider $L=3$ (so that $\ell$ can be 0,1,2), corresponding to three possible orientations of the steerable wavelet. \footnote{Note that this $\ell$ does not indicate the multipole of the spherical harmonic decomposition, as it is done in other sections of this paper, but rather the rotation index. We kept this notation in this section (and in the next one) to be consistent with the WPH literature.} The number $n$ specifies an oscillation of the order of $2^{n+1}$ pixels; as we are using patches of 128x128 pixels, $n$ runs from 0 to 5. This choice of spacing between different filter sizes follows the standard implementation of \cite{Allys2020}; for simplicity, and in order to keep our data vector size reasonably small, we chose to not explore a thinner spacing.  Note that the wavelet is real in Fourier space and is complex in real space. 

The wavelet transform of a field is the convolution of the field with ${\psi}_{n,\ell}$ (for arbitrary $n$ and $\ell$). For the  wavelet transform of the convergence map in tomographic bin $i$ we write:
\begin{equation}
\kappa_{{n,\ell}}^i(\vec{\theta}) \equiv \left(\kappa^i \ast {\psi}_{n,\ell}\right) (\vec{\theta}).
\end{equation}
Its Fourier transform for each $(n,\ell)$ has central frequency $\vec{\xi}=2^{-n}\Rot_\ell\vec{\xi_0}$ and has finite width, and thus each convolution is a local filtering in Fourier space. As shown in Fig.~2 of \cite{Allys2020}, it can identify both peaks and anisotropic filaments of different orientations. The full wavelet transform spans all of Fourier space. In addition, it has the desirable feature of being well localised in both real and Fourier space. 

Following \cite{Allys2020}, we apply a non-linear operation to the smoothed fields. The non-linear operation used is called `phase acceleration'; this operation modifies the Fourier spectrum of the smoothed field, without modifying the spatial localisation of its features.  As it is a non-linear operation, it allows us to access the non-Gaussian features of the field using second moments. Modifying the spectrum of the field, on the other hand, allows us to capture interactions between fields smoothed with different filters (and therefore different scales) that would otherwise have minimum overlapping support in Fourier space.


The smoothed and accelerated field will be called the wavelet phase harmonic. The `phase harmonic of order $q$' is defined to be
\begin{equation}
    \PH(re^{\imagunit{} \theta},q) \equiv r e^{\imagunit{} q \theta},
\end{equation}
where $r$ is the modulus of the field and $\theta$ its phase. This function leaves its input unaltered for $q=1$, and takes its modulus for $q=0$. The absolute value of the field has been shown to be a useful  non-linear operation, with the desirable property that it does not amplify noise. We consider only $q=0$ or $q=1$; although $q$ can reasonably assume other values \citep{Allys2020}, we found these other statistics did not significantly improve the constraints.

Once the fields have been transformed, we can build statistics that are second order in the input field, of the form:
\begin{equation}
    \Avg_p \Avg_{\ell} \left( \PH(\kappa_{{n_1,\ell+\Delta\ell}}^{i},{ q^{}_1}) \, \PH(\kappa_{{n_2,\ell}}^{j},{ q^{}_2}) \right).
\end{equation}
As before, we average over all pixels. We also average over all values of the rotation index $\ell$ (i.e. $0 \le \ell < L$); note that this makes sense even when $\Delta\ell \ne 0$ as the rotation indices can simply `wrap around'. These statistics are therefore functions of scales ($n_1$, $n_2$), rotation index offset ($\Delta\ell$), phase harmonic orders ($q_1$, $q_2$), and tomographic bins ($i$, $j$). The statistics used in this work are:
\begin{equation}
    S00(i, j, n) =  \Avg_p \Avg_{\ell} \left( |\kappa_{{n,\ell}}^i| \, |\kappa_{{n,\ell}}^j|\right)
\end{equation}
\begin{equation}
    S11(i, j, n) \equiv \WPHG(i, j, n) = \Avg_p \Avg_{\ell} \left( \kappa_{{n,\ell}}^i \, \kappa_{{n,\ell}}^j \right)
\end{equation}
\begin{equation}
    S01(i, j, n) = \Avg_p \Avg_{\ell} \left( |\kappa_{{n,\ell}}^i| \,\, \kappa_{{n,\ell}}^j \right)
\end{equation}
\begin{equation}
    C01{\delta \ell 0}(i, j, n_1, n_2) =  \Avg_p \Avg_{\ell} \left( |\kappa_{{n_1,\ell}}^i| \,\, \kappa_{{n_2,\ell}}^j \right) \textrm{ for } n_1<n_2
\end{equation}
\begin{equation}
    C01{\delta \ell 1}(i, j, n_1, n_2) = \Avg_p \Avg_{\ell} \left( |\kappa_{{n_1,\ell+1}}^i| \,\, \kappa_{{n_2,\ell}}^j \right) \textrm{ for } n_1<n_2.
\end{equation}
%
%
Here $i,j$ vary over the four tomographic bins, whereas $n$ (or $n_1$ and $n_2$) varies over the possible wavelets under consideration. Following \cite{Allys2020} we use `$S$' for the statistics with $n_1=n_2=n$ ($S$00, $S$11, and $S$01) and `$C$' for the statistics with $n_1<n_2$ ($C01{\delta \ell 0}$ and $C01{\delta \ell 1}$) that capture correlations at different wavelet scales. 

The statistics probe non-Gaussian features of the field (with the  exception of $S11$, which is Gaussian in that it is equivalent to the power spectrum of $\kappa$; for this reason we refer to it as `$\WPHG$'). One advantage of the WPHs over conventional moments is that they are always `second-order' in the input field, which makes them more robust against additive noise \citep{Allys2020}. Additional statistics using more combinations of WPHs could have been considered, as in \cite{Allys2020}; however, for computational reasons we restrict ourselves to the summary statistics listed (having checked that they capture nearly all the information given the noise levels in our data). In total, we have 60 components for $S11$ (ten independent tomographic bin pairs and six scales), 96 components for $S00$ and $S01$ each (16 tomographic bin pairs and six scales), and 240 components for $C01\delta \ell0$ and $C01\delta \ell1$ each (16 tomographic bin pairs and 15 scale pairs).

As in the case of moments, we subtract some specific WPH moments of noise-only maps from our estimators. In particular, for WPH $S01$, $C01{\delta \ell 0}$, and $C01{\delta \ell 1}$ we subtract a term involving one noise-only map and the observed noisy convergence map. We empirically found these statistics to be the ones mostly affected by source clustering, and this subtraction to be the best way to minimise source clustering effects.

\subsection{Scattering Transform}\label{sect:ST}

The scattering transform \citep{Mallat2012,Bruna2013,Cheng2020,Valogiannis2022a,Valogiannis2022} is in concept similar to the WPHs introduced above. The idea is to smooth the field using the directional, multi-scale wavelet transform, followed by a modulus operation on the field. This pair of operations can then be reapplied several times; we finish with an overall average over the sky. This yields a hierarchy of scattering transform coefficients $\ST_{m}$, where $m$ is the number of smoothing and modulus operations applied. This work uses scattering coefficients of order $m=1,2$. Given a directional multi-scale wavelet $\psi_{n,\ell}$ and the convergence map $\kappa^i$ of tomographic bin $i$, we obtain:
\begin{equation}
    \ST_{1}(i, n) = \Avg_p \Avg_{\ell} \left( | \kappa^i  \ast \psi_{n,\ell} | \right)
\end{equation}
\begin{multline}
    \ST_{2}(i, n_1, n_2, \ell') =  \Avg_p \Avg_{\ell} \left( | \, | \kappa^i  \ast \psi_{n_1,\ell} | \ast \, \psi_{n_2,\ell'-\ell} | \right)\\ \textrm{ for } n_1 \le n_2.
\end{multline}
The average over $\ell$ in $\ST_{2}$ makes this summary statistic invariant to rotation, while preserving morphological information. The $\ST_1$ coefficients are qualitatively similar to the power spectrum amplitudes, weighted by a window function, but while the power spectrum uses the $L^2$ norm of the convolved field, the scattering transform uses the $L^1$ norm. 
The $\ST_2$ coefficients probe more non-Gaussian information stored in the field, providing the co-occurrence information at the scales $n_1$ and $n_2$ and capturing interferences of the field between features selected with two successive wavelets. 

The scattering transform coefficients are `first-order' in the input field. To enable the computation of scattering transform coefficients including pairs of maps of different tomographic bins, we follow \cite{Zuercher2021b} and introduce the `cross-maps' $\kappa^{ij}(\theta, \phi)$ 
\begin{equation}
    \label{eq:cross_maps}
    \kappa^{ij}(\theta, \phi) = \sum_{\ell = 0}^{\ell_{\mathrm{max}}}\sum_{m = -\ell}^{\ell} \hat{\kappa}^i_{ \ell m} \hat{\kappa}^j_{\ell m} Y^{}_{\ell m}(\theta, \phi),
\end{equation}
where $i$ and $j$ (with $i>j$) denote two different tomographic bins. We then compute the scattering coefficients $\ST_1$ and $\ST_2$ of the cross-maps. In total, we consider 60 coefficients for $\ST_1$ (six scales, ten independent tomographic bins), and 630 for $\ST_2$ (21 scale combinations, ten tomographic bins, and three different orientations). 

{The ST is similar to the WPHs, but with a few differences. First, the scattering transform stays `first-order’ in the observed field, whereas the WPHs are always `second-order'. This means that the ST is less susceptible to noise than the WPHs. Second, in the WPHs there is a natural definition of cross-correlation between different fields; this is not the case for the ST (it is for this reason that we introduced the cross-maps so as to account for cross-correlations between different tomographic bins). As we will see in \S \ref{sect:end-to-end}, this has an impact on the constraints, as the ST deals with cross-correlations less efficiently. This also applies to cross-correlations between different scales for the non-Gaussian features: the WPHs use the cross correlations in combination with the phase acceleration as a non-linear operation to couple scales (e.g. the WPH C01 coefficients); an analogous statistic cannot be defined for the ST.}

{ST and WPH are often compared to machine learning methods as they were designed to emulate information capture in the manner of a convolutional neural network (CNN), without the need for training data. This is quite evident especially for the ST coefficients: the smoothing of the field is equivalent to the CNN kernel convolution, the modulus operation is equivalent to the CNN ReLU layer, the average is equivalent to the CNN `pooling', and the hierarchy of coefficients is equivalent to the CNN's multiple layers. The analogy, however, stops here: for the ST and WPH, since there is no training, we have full control over the kernels (i.e. the wavelets), or on all the details of the summary statistics (i.e. the order of the phase acceleration for the WPH, or how different tomographic bins are combined). This is different to CNNs, which are commonly referred to as `black boxes' because of the difficulties associated with comprehending the features they learn and the significance of the numerous parameters acquired during training.}

\subsection{Multipole support, covariance, and signal-to-noise of the summary statistics}
\begin{figure*}
\includegraphics[width=0.9 \textwidth]{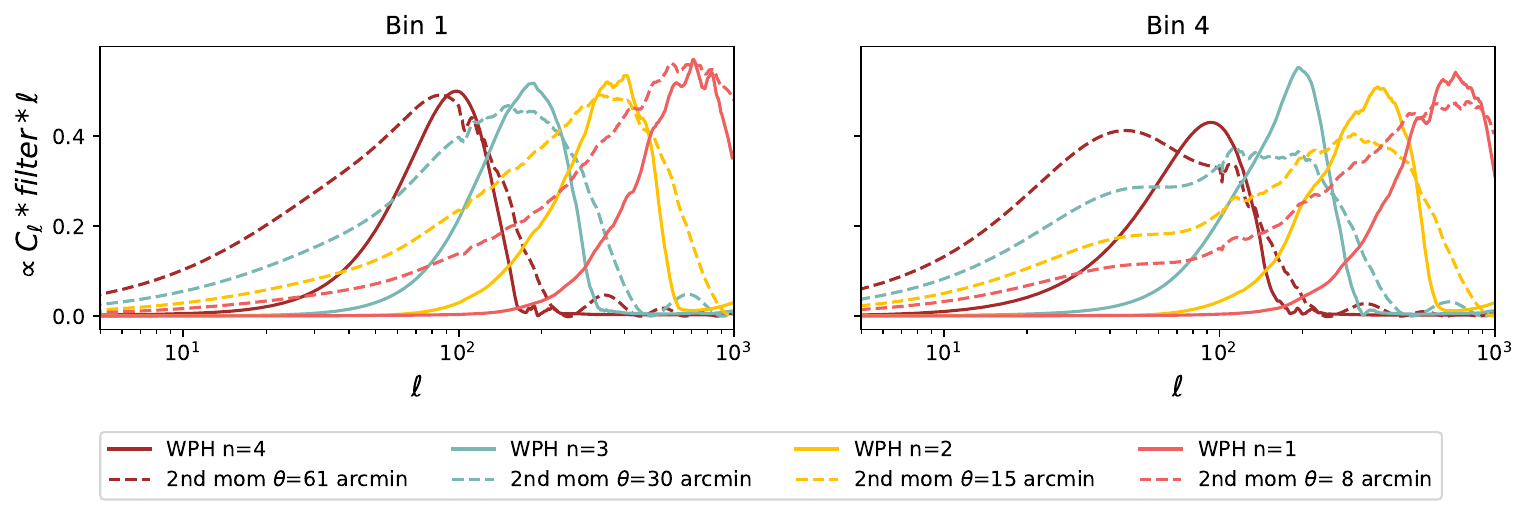}
\caption{Power spectra ($C_{\ell}$) of convergence maps of the first (left) and last (right) tomographic bins; the maps have been smoothed by directional wavelet (solid line) and top hat (dashed line) filters of different sizes. For plotting purposes the measured power spectra have been smoothed with a Savitzky–Golay filter.}
\label{fig:filter_support}
\end{figure*}
\begin{figure}
\includegraphics[width=0.45\textwidth]{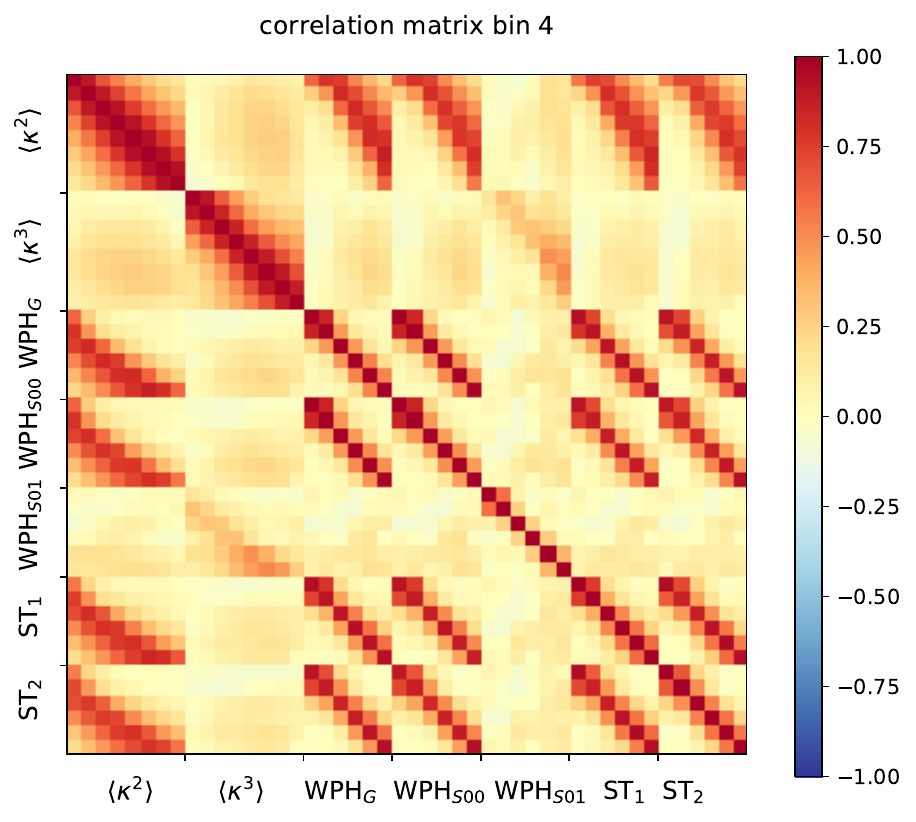}
\caption{Correlation matrix for some of the statistics considered in this work, computed from the \texttt{CosmoGridV1} simulations at the fiducial cosmology. We consider only the fourth tomographic bin.}
\label{fig:corr}
\end{figure}

\begin{figure}
\includegraphics[width=0.45\textwidth]{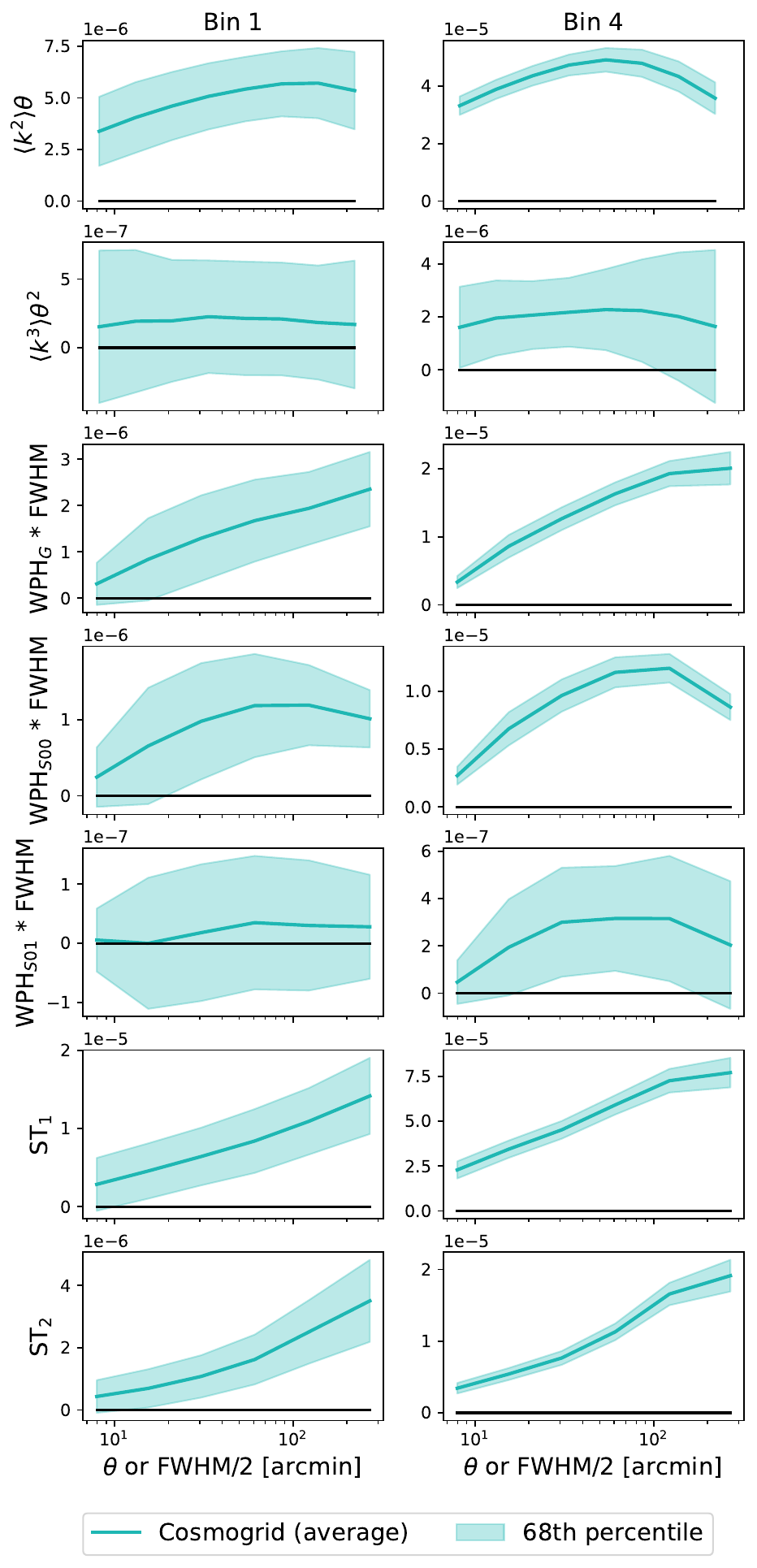}
\caption{Some of the Gaussian and non-Gaussian statistics considered in this work, as measured in simulations. The left (resp. right) column shows the statistics measured only using the noisy convergence map of the first (resp. fourth) tomographic bin. The calculation was done for multiple \texttt{CosmoGridV1} simulations at the fiducial cosmology; solid lines are the resulting average and the bands are 68th percentiles. For statistics involving filters with different sizes $j_1,j_2$, we considered $j_1=j_2$.}
\label{fig:DV}
\end{figure}

The statistics considered here implement different filters to smooth the convergence field, so it is instructive first to look at the support in multipole space covered by the smoothed maps. This is shown in Fig.~\ref{fig:filter_support}, which plots (for the first and last tomographic bin convergence maps) the power spectra of the smoothed maps, where this smoothing is done using top-hat and directional wavelet filters of different sizes. To compare roughly the two types of filter, the top hat filter radii $\theta$ have been chosen to be half the FWHM of the wavelet filters. Fig.~\ref{fig:filter_support} shows that the maps smoothed by the two sets of filters peak roughly at the same point in multipole space, but that the top hat filters are much less localised. This is expected as wavelets are designed to better isolate scales, both in real and in multipole space.

The statistics considered here are also in part covarying, i.e. they probe similar information. Therefore it is instructive to construct the correlation matrix of the data vector; this can be done starting from the $400$ mock measurement of the \texttt{CosmoGridV1} simulations and by computing the covariance matrix:
\begin{equation}
\label{eq:cov_def}
\hat{{C}} = \frac{1}{\rm N_s}{\sum_{i=1}^{\rm N_s}} (\hat{{d}}_i-\hat{{d}}){(\hat{{d}}_i-\hat{{d}})}^{T},
\end{equation}
where ${\rm N_s}$ is the number of simulations, $\hat{{d}}_i$ the data vector measured in the $i$-th simulation, and $\hat{{d}}$ the sample mean.  The elements of the correlation matrix $\widehat{{\rm Corr}}$ can be obtained as
\begin{equation}
\widehat{{\rm Corr}}_{i,j} = \frac{\hat{{\rm C}}_{i,j}}{\sqrt{\hat{{\rm C}}_{i,i}\hat{{\rm C}}_{j,j}}}.
\end{equation}
This is illustrated in Fig.~\ref{fig:corr}, which shows the correlation matrix of the different statistics as a function of scales. For the sake of simplicity, we considered only the part of the data vector including the fourth tomographic bin. We make three remarks:
\begin{itemize}
\item{Second and third moments blocks are much more correlated than those of the scattering transform and the WPH. This is a consequence of the smoothing filter adopted: wavelet filters are significantly better at isolating scales, and this makes the correlation matrix more diagonal.}
\item{Gaussian statistics (second moments and $\WPHG$) are highly correlated, as expected. They are also highly correlated with WPH S00 and the scattering coefficients $\ST_1$ and $\ST_2$. The latter are probing both Gaussian and non-Gaussian features of the field, although this figure suggests they weigh Gaussian features more.}
\item{Third moments and WPH $S01$ are not very correlated with their Gaussian counterparts (a fact also exacerbated by shape noise), but they are mildly correlated with each other. This suggests WPH $S01$ is in part probing the bispectrum of the field. Although not shown in the figure, we report that WPH $C01\delta \ell0$ and $C01\delta \ell1$ behave similarly to WPH $S01$.}
\end{itemize}

{We report the signal-to-noise ratio (SN) of the different statistics in Table~\ref{scalesetc}. We note that this SN is computed for the part of the measurements that only uses one tomographic bin.} The Gaussian statistics considered in this work have significantly higher signal-to-noise compared to third moments or WPH $S01$. On the other hand, WPH $S00$, $\ST_1$ and $\ST_2$ have significance similar to Gaussian statistics, as they are also probing Gaussian information of the field. Among the purely non-Gaussian statistics, we note that WPH $S01$ has a higher signal-to-noise ratio than that of the third moment. This is due to the former statistic being only `second-order' in the input field, which makes it less affected by noise.  

Last, we show in Fig.~\ref{fig:DV} some of the statistics as measured in \texttt{CosmoGridV1} simulations at the fiducial cosmology.


\begin{table*}
\caption {Salient properties of the summary statistics. The second column denotes whether it carries Gaussian (G) or non-Gaussian (NG) information. The third column refers to the order of the field $\kappa$. The fourth column is the number of components of the datavector across scales and tomographic bins. The further columns show the signal-to-noise ratio (SN) of the measurements in the \texttt{CosmoGridV1} simulations for each tomographic bin. We note that this is not the total SN of the full measurement, but only the SN of the measurement for one tomographic bin.}
\centering
\begin{tabular}{c|c|c|c|c|c|c|c}
 \hline
& \textbf{G/NG} & \textbf{Order} & \textbf{Length of Datavector} & \textbf{Bin 1 SN} & \textbf{Bin 2 SN} & \textbf{Bin 3 SN} & \textbf{Bin 4 SN} \\
 \hline
2nd moments & G & 2  &  160     & 3.4 & 7.8 & 16.1 & 15.2\\
3rd moments & NG & 3 &  512    & 0.8 & 0.9 & 1.7 & 1.3\\
WPH S11 ($\WPHG$) & G & 2 & 120   & 3.1 & 7.4 & 15.6 & 14.4\\
WPH S00 & NG & 2 & 96        & 2.8 & 6.9 & 14.9 & 13.6\\
WPH S01 & NG & 2 & 480        & 0.7 & 1.5 & 2.9 & 2.4\\
ST1  & NG & 1 & 60         & 3.3 & 7.8 & 15.3 & 15.3\\
ST2  & NG & 1 & 630         & 3.1 & 7.3 & 15.3 & 15.3\\
\hline
\end{tabular}
\label{scalesetc}
\end{table*}


\subsection{Data Compression}

\begin{figure}
\includegraphics[width=0.45 \textwidth]{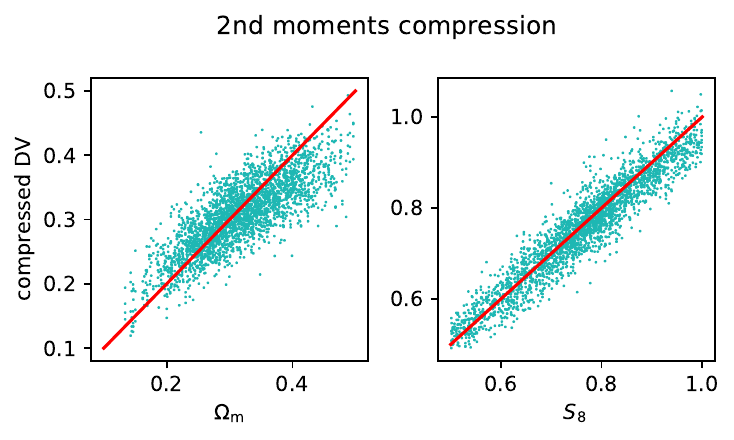}
\includegraphics[width=0.45 \textwidth]{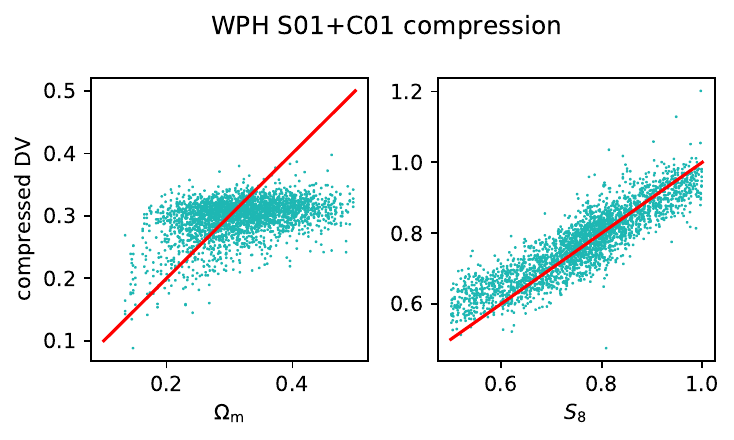}
\caption{Example of compressed statistics using second moments (top) and PWH S01+C01 (bottom) as summary statistics,  and $\Omega_{\rm m}$ and $S_8$ for target parameters for the loss function. The y-axis is the compressed statistic, while the x-axis is the true value of the parameter. Each point represents one input measurement. The red line serves to guide the eye and indicates an unbiased compression. Generally, the tighter the scatter, the better the given
statistic is at constraining that parameter; on the contrary, a broader and biased compression indicates poor sensitivity to that parameter (e.g. WPH S01+C0 for $\Omega_{\rm}$). A biased compression does not imply a biased inference, as we compress the simulated measurements and the data in the same way.}
\label{fig:compression_example}
\end{figure}

Data compression is paramount in the likelihood-free inference framework, as for a fixed number of simulated mocks the density estimation is more efficient when the dimensionality of the data vector is low \citep{jeffrey_lfi}. Different compression methods exist (e.g. PCA-based compression, \citealt{Zuercher2021}; MOPED, \citealt{Heavens2000}; neural compression, \citealt{jeffrey_lfi}). Notably, a poor compression scheme could result in less informative summaries, but it would not produce biased results. For this work we follow \citealt{jeffrey_lfi} and use a neural compression scheme to compress the summary statistics to the same dimension as the parameters $\theta$ in which we are interested (but see Appendix \ref{sect:MOPED} for a comparison with the MOPED compression). In particular, given a summary statistic $\mathbf{d}$, we compress it using $\mathbf{t} = F_{\phi}(\mathbf{d})$, and we approximate $F_{\phi}$ by a neural network. We determine the neural network parameter $\phi$ by minimising a Mean Squared Error (MSE) loss function using the first half (12656) of our pseudo-independent mocks. The architecture used for the network and the number of parameters are summarised in Table~\ref{tab:layers}.  Since in this work we consider multiple summary statistics and their combinations, we chose to compress summary statistics individually and to combine their compressed versions (i.e. stack the data vectors) later on during the likelihood-free inference process. In particular, we individually compress second moments, third moments, $\WPHG$, WPH $S00$, $\ST_1$, and $\ST_2$. The only exception is for WPH $S01$ and WPH $C01$, which are compressed together. We compress the data vectors using all the parameters, one at a time. Examples of compression are shown in Fig.~\ref{fig:compression_example} for second moments and WPH S01+C01, against the parameters $\Omega_{\rm m}$ and $S_8$. Generally, the tighter the scatter, the better the given statistic is at constraining that parameter. For second moments, the compressed statistics trace fairly well the parameter against which they have been compressed; on the other hand,  the WPH S01+C01 case shows a poor sensitivity to $\Omega_{\rm m}$. The compression is not expected to be `unbiased': as it can be seen from Fig.~\ref{fig:compression_example}, the compressed statistics do not recover the true value of the simulations (the red line in the plot), even in the best case ($S_8$ for second moments). This is not a problem for the inference; as we consistently compress both the data vectors measured in simulations and the data, the final posterior will be unbiased.

\begin{table}
\centering
\caption{Neural Network Layers and number of parameters used for the compression of the summary statistics.}
\label{tab:layers}
\begin{tabular}{c|c|c}
\hline
\textbf{Layer (type)} & \textbf{Output Shape} & \textbf{Number of Parameters } \\ \hline
Dense &  900 & 900*(length DV+1)   \\ 
LeakyReLU & 900 & 0 \\ 
Dense & 800 & 720800     \\ 
LeakyReLU &  800 & 0 \\ 
Dense &  100 & 80100    \\ 
ReLU &  100 & 0 \\ 
Dense & 100 & 10100   \\ 
ReLU &  100 & 0 \\ 
Dense & 1 & 101 \\ \hline
\end{tabular}
\end{table}


\section{Likelihood-free inference}\label{sect:LFI}

In likelihood-free inference (also known as simulation-based inference), the likelihood $p(d|\theta)$ is not assumed to have a closed form; rather, it is reconstructed from simulated mock data as part of the inference pipeline. Here is a summary of the procedure used to infer the posterior distribution of the parameters; a more detailed description is provided in \cite{jeffrey_lfi}.

In our implementation, the parameter inference task is posed as a density estimation problem. Let us assume we have a set of mock noisy data vectors $d$ and simulation parameters $\theta$ forming a cloud of points in $\{d, \theta \}$ space. We then estimate the conditional distribution $p(d|\theta)$ with an ensemble of neural density estimators (NDEs): specifically, we use both Gaussian Mixture Density Networks (MDN; \cite{Bishop1994})  and Masked Autoregressive Flows (MAF; \cite{Papamakarios2017}).  We used two MDNs with two and three Gaussian components respectively, each with two dense hidden layers with 30 neurons per layer, and we used two MAFs with two and three MADE (Masked Autoencoders for Distribution Estimation, \cite{Germain2015}) layers respectively, each with two dense hidden layers with 50 neurons per layer.  For each of our neural density estimation methods, MDN and MAF, the network was trained to give an estimate $q(d|\theta; \phi)$ of the target distribution $p(d|\theta)$ (i.e. $p(d|\theta) \approx q(d|\theta; \phi)$); here $\phi$ are the parameters of the network, determined by minimising a loss function $U(\phi) = - \sum_{n=1}^{N} {\rm log} q(d_n|\theta_n; \phi)$ over the N forward-modelled mock data $d_n$.  This loss corresponds to minimizing the Kullback-Leibler divergence, a measure of difference or change going from the estimate $q$ to the target $p(d|\theta)$. To perform the density estimation and the training we used the publicly available package pyDELFI \citep{Alsing2018}.

The final density estimation is a stack of the ensemble estimates, weighted by the loss evaluated during training. Once the target distribution $p(d|\theta)$ has been estimated, we evaluate it at the observed data $d=d_{\rm obs}$ to obtain the likelihood. For completeness, we show in Appendix \ref{sect:indivuals} the posteriors obtained by each NDE and how they differ from the stacked posterior.

{Using NDEs to infer the \textit{likelihood} surface rather than the posterior has one main advantage: as long as the parameters varied in the simulations are taken into account during the training process, the fact that the parameter space is not sampled uniformly does not translate into an effective prior on our final constraints, i.e. it does not produce tighter posteriors \citep{Alsing2018}. This means that after we trained the NDEs and learned the likelihood surface, we can use a different prior during the inference when estimating our posteriors (see Table~\ref{parameter} for the priors used in the analysis). Of course, in the regions of the parameter space where we only have a few simulations, the estimation of the likelihood surface will be noisier and the likelihood less accurate; this is why the Gower St simulations have been run in active-learning mode for $\Omega_{\rm}$ and $S_8$, to increase the accuracy of the likelihood estimation in the region covered by the data posterior.}

{For practical reasons, due to our limited number of mocks, it is not possible to reliably estimate the likelihood surface taking into account all the parameters varied in the simulations. As we are mostly interested in the constraints on $\Omega_{\rm m}$, $S_8$, $w$, and $A_{\rm IA}$, the main density estimation was carried out using the parameters $\theta = \left[ \Omega_{\rm m}, S_8, w, A_{\rm IA}\right]$ and the associated compressed data vectors. This means that the other parameters are effectively marginalised over; this time, however, since we are not explicitly taking into account their dependence during the training of the NDEs, the parameter distribution \textit{does} matter. This is explained via \textit{marginal posterior density} estimation in ~\cite{momentnets};  we can therefore assume their marginalisation follows the prior distribution used to sample these parameters when generating the mocks as reported in Table~\ref{parameter}. }

To train the NDEs we used the compressed data vectors and mocks that were not used to train the compression algorithm (i.e. the last 12656 pseudo-independent mocks). Whenever we combine different summary statistics, we stack the individual compressed data vectors together.  We restrict the density estimation procedure to our eventual prior range (Table~\ref{parameter}). The final posteriors are then obtained through Markov chain Monte Carlo (MCMC) sampling of the likelihood, assuming the priors listed in Table~\ref{parameter}. The MCMC sampling is performed using the public software package \texttt{EMCEE} \citep{Foreman-Mackey2013}, an affine-invariant ensemble sampler for MCMC.

\begin{figure}
\includegraphics[width=0.4 \textwidth]{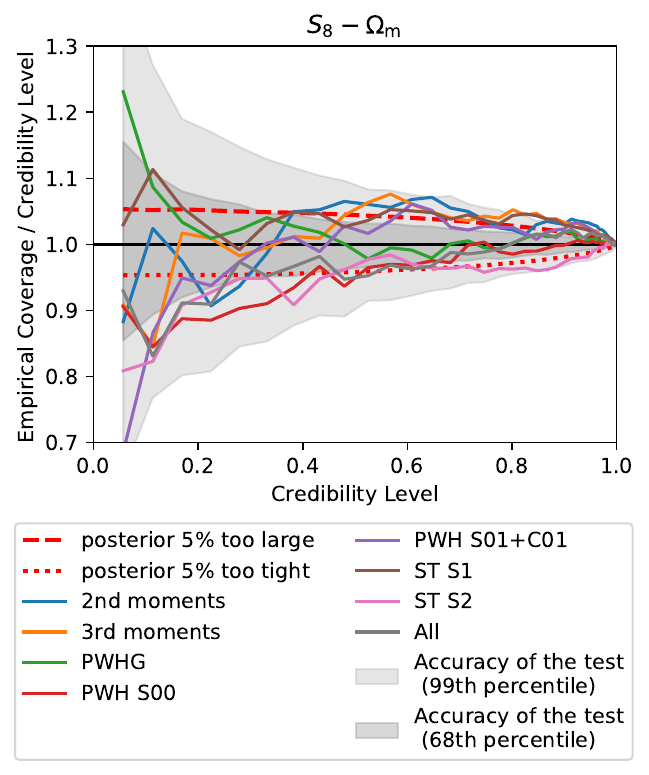}
\includegraphics[width=0.4 \textwidth]{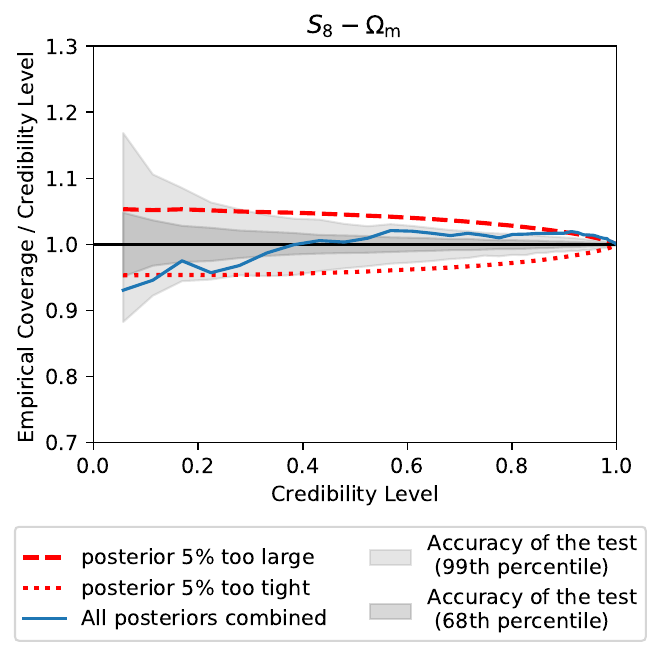}
\caption{Expected coverage probability of the posteriors obtained using the LFI pipeline and different summary statistics with respect to the credibility level. The two red dashed/dotted lines indicate what the expected coverage probability would be if the posteriors were misestimated by 5 per cent. The grey shaded regions indicate the accuracy of the test given the limited number of posteriors (500) used here. The top panel shows the test for each of the summary statistics considered in this work and their combination; the bottom panel uses all the posteriors (500x8=4000) to test the size of the posteriors with a higher accuracy. }
\label{fig:cumulative_posterior}
\end{figure}

To test that the confidence levels obtained through the likelihood-free-inference are not misestimated, we perform an empirical coverage test. {We first select a subset (125) of the full-sky Gower St simulations uniformly spanning the $\Omega_{\rm m}-S_8-w$ space.  We do this by uniformly dividing each dimension into 5 parts, so as to partition the three-dimensional space into 5x5x5 cuboids, and by selecting only one simulation per cuboid. For this test, we excluded the outermost regions of our parameter space close to the edge of the priors, where we know we only have a few simulations and the likelihood estimation might be uncertain: in particular, we only selected simulations in the range $\Omega_{\rm m} \in [0.2,0.4]$, $\S_{8} \in [0.6,0.9]$, and $w \in [-1,-0.5]$. For each of the full-sky simulations, we choose four non-overlapping DES Y3 mocks (picked at random from the different noise realisations), for a total of 500 mocks. We re-train our compression algorithm and NDEs excluding these mocks; then,  we obtain posteriors for each of them and check the confidence regions that cover the true values of $\Omega_{\rm m}$ and $S_8$.} Finally, we report in Fig.~\ref{fig:cumulative_posterior} the fraction of posteriors encompassing the true value at a given confidence level. A perfectly calibrated posterior would have an expected coverage probability equal to the credibility level. Overconfident posteriors (i.e. tighter than they should be) would lie in the bottom right part of the plots; on the other hand, conservative posteriors (i.e. larger than they should be) would lie in the upper left part of the plot. The number of posteriors we ran limits the accuracy of this test; with 500 posteriors per summary statistics, we can determine if the posterior size is accurate at the $\sim$ 5 percent level. The statistics considered are consistent with a perfect calibrated posterior within the accuracy of the test. When all the posteriors are considered (lower panel of Fig.~\ref{fig:cumulative_posterior}), the scatter reduces significantly, indicating no bias in the size of the posterior at the level of a few percent.

In Appendix \ref{sect:likelihood_tests} we provide further tests concerning the NDE likelihood estimates using \texttt{CosmoGridV1} simulations.

\subsection{Comparison between approaches with theory-based models and Gaussian likelihood}\label{sect:gaussian_likelihood_}
\begin{figure}
\includegraphics[width=0.45 \textwidth]{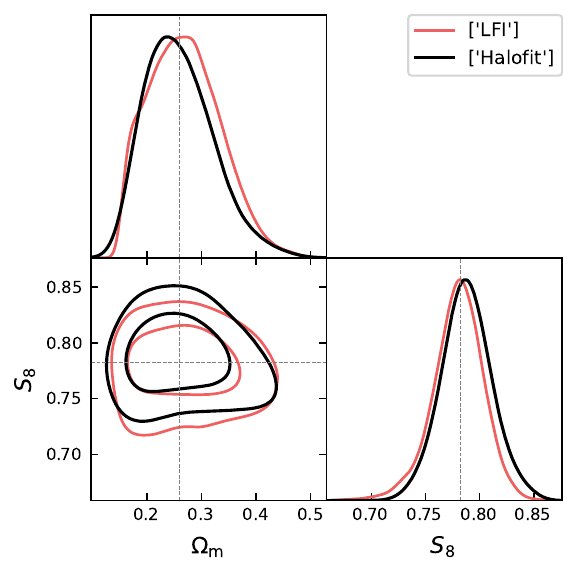}
\caption{Posteriors for $S_8$ and $\Omega_{\textrm{m}}$ obtained by analysing the power spectrum of one of the Gower St simulations with two pipelines: the LFI pipeline described in this work and a different one that uses the theory model described in \protect\cite{Doux2022} and that assumes a Gaussian likelihood.}
\label{fig:halofit_vs_cosmosis}
\end{figure}

We perform in this section a comparison between a) the cosmological constraints obtained using the LFI pipeline and b) a more standard approach in which we rely on a theoretical model for the observables and we assume the likelihood to be Gaussian. To this end, we use as a summary statistic the (pseudo) power spectrum, as implemented in \cite{Doux2022}. Most of the summary statistics explored in this work do not have a theoretical model, except for the second and third moments \citep{G20,moments2021}; the code available to us to model moments, however, does not allow us to marginalise over the neutrino mass or over $w$. Moreover, we do not have a theoretical model for the covariance, which is, on the contrary, available for the pseudo power spectrum analysis. For these reasons we decided to use the power spectrum as a summary statistic for this comparison.

To perform the comparison, we analysed a theory data vector at a fiducial cosmology. As a minor caveat, we created (specifically for this test) mocks without source clustering (i.e. we assumed $b_g=0$ and $F(p) = 1$ in Eq.~\ref{eq:sc_pixel}), as this effect is not included in the theory model for the power spectrum; moreover, with source clustering the noise is slightly cosmology dependent, and this effect is not captured by the theory covariance implemented in \cite{Doux2022}. Without source clustering, we note that the IA model reduces to a pure NLA model.

We then analysed the same noisy data vector using the theory model of \cite{Doux2022}, which is based on halofit \citep{Takahashi2012}. We sampled the posteriors of our parameters using \textsc{Polychord} \citep{poly1,poly2}; this is a nested sampler that uses slice sampling within the nested iso-likelihood contours. For the cosmological parameters, we varied the same parameters spanned by our mocks (see Table~\ref{parameter}), and, where possible, we assumed the same priors.  For $\Omega_{\rm b}$, $h_{\rm 100}$, $n_{\rm s}$, and neutrinos, we assumed flat priors, but we later importance-sampled the posterior to reflect the Gower St effective priors. 

The posteriors for $S_8$ and $\Omega_{\rm m}$ from the two pipelines are shown in Fig.~\ref{fig:halofit_vs_cosmosis}, showing an excellent agreement. This agreement is not trivial: it relies on the validity of the Gaussian likelihood assumption for the power spectrum analysis, on the forward modelling of our simulations to be equivalent to the modelling used by the theory pipeline of \cite{Doux2022}, and on the dependence of the covariance on cosmological and nuisance parameters to be negligible. In other words, a lack of agreement would not have invalidated our pipeline; rather, it would have challenged some of the main assumptions behind standard Gaussian likelihood analyses of Gaussian statistics such as found in \cite{Doux2022}; \cite{y3-cosmicshear1}; \cite*{y3-cosmicshear2}. {The primary validation tests for assessing the accuracy of our posterior estimates include the empirical coverage test outlined in the preceding section, as well as the scale-cut tests and the end-to-end pipeline test discussed in the subsequent sections (\S \ref{sect:scale_cuts} and \S \ref{sect:end-to-end}).}

\section{Scale cuts}\label{sect:scale_cuts}

\begin{figure*}
\begin{center}
\includegraphics[width=0.9 \textwidth]{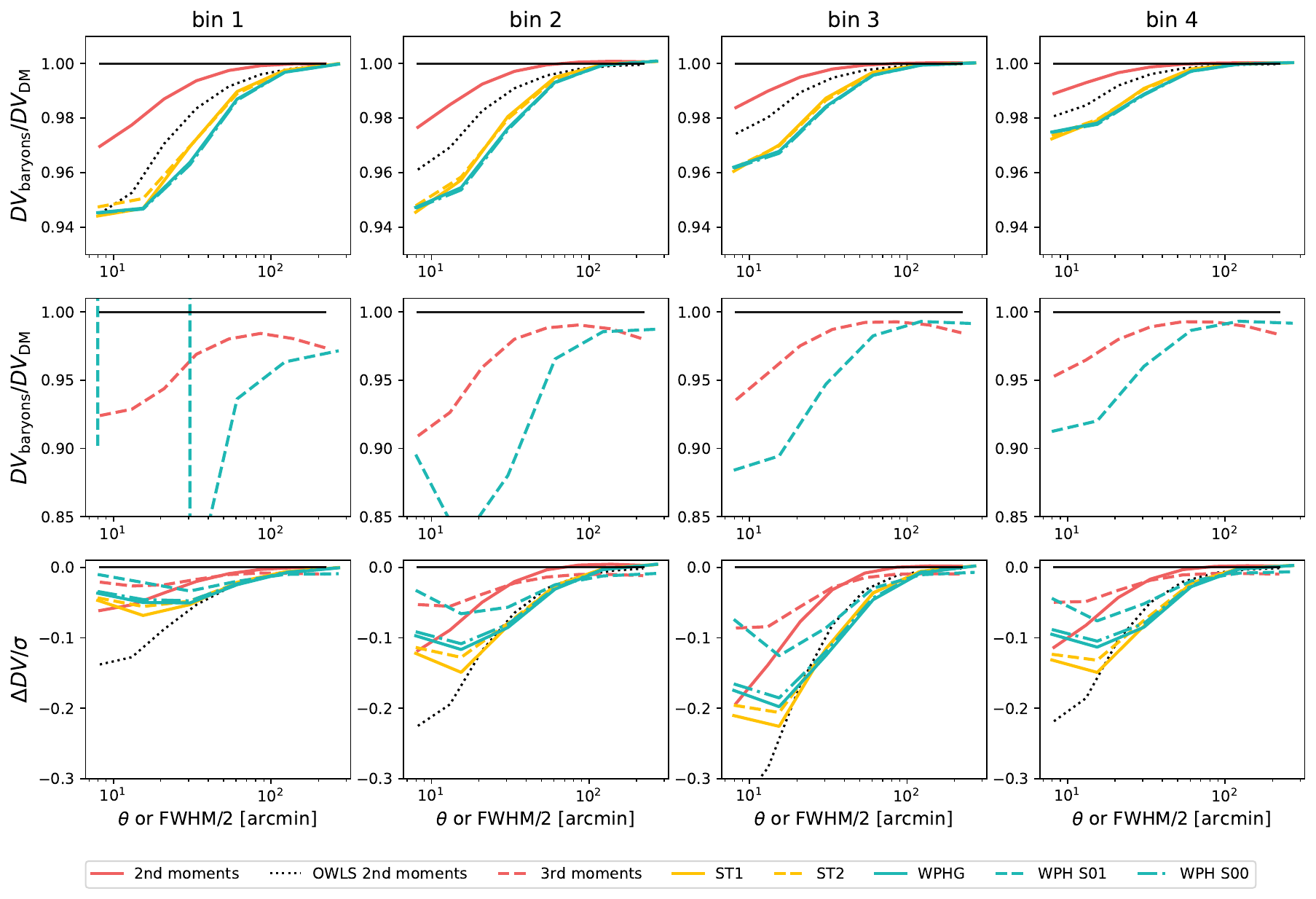}
\end{center}
\caption{Impact of baryonic feedback effects on the summary statistics considered in this work. Each column shows the impact on the part of the summary statistics obtained using maps from a specific tomographic bin. The top two rows show the ratio between the  data vector as computed in simulations with and without baryonic feedback; the bottom row shows the difference between the data vectors normalised by the square root of the diagonal of the covariance matrix. For the second moment, we also show the result for the OWLS-AGN simulations (dotted line as indicated in the legend). We note that in the first bin, the ratio between the WPH S01 data vector with and without baryonic contamination changes sign at small scales; this is because at those scales the amplitude of the data vector changes sign and it is close to zero.}
\label{fig:effectofbaryons}
\end{figure*}

\begin{table}
\caption{Bias of the parameter posteriors assessed by comparing the outcomes of an analysis performed on a simulation with baryonic feedback to that of a simulation without baryonic feedback. Biases for different summary statistics are reported in terms of the distance between the peaks of the posteriors in the $S_8-\Omega_{\rm m}$ plane. All the biases are smaller than $0.3\sigma$ (the maximum level of bias accepted by our analysis).}
\centering
\begin{tabular}{c|c}
 \hline \textbf{Summary Statistic(s)}
& \textbf{Contamination} \textbf{$S_8-\Omega_{\rm m}$}\\
 \hline
2nd moments & 0.01$\sigma$ \\
$\WPHG$  & 0.05$\sigma$ \\
3rd moments & 0.09$\sigma$ \\
WPH S00 & 0.01$\sigma$ \\
WPH S01+C01 & 0.04$\sigma$ \\
WPH S00+S01+C01 & 0.11$\sigma$ \\
ST1 & 0.03$\sigma$ \\
ST2 & 0.03$\sigma$ \\
ST1+ST2 & 0.06$\sigma$ \\
2nd+3rd moments & 0.03$\sigma$ \\
2nd moments+WPH & 0.03$\sigma$ \\
2nd moments+ST1 & 0.05$\sigma$ \\
2nd+3rd moments+ST+WPH & 0.05$\sigma$ \\
\hline
\end{tabular}
\label{fig:contamination}
\end{table}

%

We determine in this section if we need to remove scales from our analysis because of a lack of modelling and/or potential systematic contamination. We test three main effects: 1) baryonic feedback processes; 2) additive biases due to PSF errors; 3) residual source clustering contamination. To anticipate the results of this section, we state here that we found all these effects to be negligible; therefore, our main analysis retains all the scales considered so far.

\subsection{Impact of lack of modelling Baryonic feedback}

The main limitation of our analysis is the lack of a proper model for baryonic feedback processes at small scales. The modelling of our observables relies on our ability to produce realistic mock catalogues; at small scales, this requires an ability to contaminate the mock catalogues with a variety of baryonic feedback models. Tools to create such contaminated catalogues exist; for example, baryonic correction models \citep{Schneider2015,Arico2020} can adjust the particle positions in gravity-only simulations to mimic the impact of various baryonic processes on the density distribution. These models have been shown to be flexible enough to accurately replicate the 2-point and 3-point statistics of various hydrodynamical simulations. Unfortunately, the simulations we use for this project have not been post-processed with the baryonic correction model, forcing us to remove scales that can be potentially affected by baryons. This is also in line with the main DES Y3 strategy for weak lensing analyses (e.g. \citealt{y3-cosmicshear1};\citealt*{y3-cosmicshear2}; \citealt{moments2021,Zuercher2021}), which did not attempt to model baryonic processes but rather removed scales potentially affected by them.

To determine which scales to remove, we use another set of public gravity-only simulations (\texttt{CosmoGridV1}) that have been post-processed with the baryonic correction model. For each full-sky simulation (with and without the baryonic correction model), we cut out four DES Y3 footprints and produce ten different noise realisations using our pipeline, totalling to two sets of 400 DES Y3 mock catalogues. The impact of the baryonic feedback model on the statistics used in this work is shown in Fig.~\ref{fig:effectofbaryons}. 

The main effect of the baryonic model adopted is to suppress the values of the measured statistics, at all scales, with more dramatic effects in the first tomographic bin (first and second rows of Fig.~\ref{fig:effectofbaryons}). 
Statistics based on wavelets seem to be affected more by baryonic feedback than moments, as the latter rely on top-hat smoothing. This is, however, not a real problem, as it is due to the top hat filters being broader and skewed towards smaller multipoles / larger scales, not affected by baryons; this dilutes the baryonic contamination. For practical purposes it is actually better to have filters with a more compact support, as this makes it easier to remove the part of the measurements affected by systematics.

The impact of baryons on non-Gaussian statistics can be qualitatively different from their Gaussian counterparts \citep{Foreman2021,Arico2020}. In a first approximation a suppression of the underlying density field should translate into a suppression of N-point statistics that will be larger as the order of the statistics increases. Fig.~\ref{fig:effectofbaryons} indeed shows a larger impact of baryons on the amplitude of the data vector for third moments compared to second moments. For the other non-Gaussian statistics included in this work, however, it is more difficult to apply this qualitative argument: ST and WPH are either linear or second order in the input field, and many of them are highly correlated with the Gaussian statistics. The impact of baryons on the amplitude of WPH S01 is significantly larger compared to $\WPHG$, but for all the other non-Gaussian statistics, the suppression is basically the same as that of $\WPHG$. 




To determine which scales to remove from our analysis, we check that the posterior on the cosmological parameters obtained by analysing a data vector from the simulations with baryonic feedback is not substantially biased with respect to the posterior obtained from a data vector measured in simulations without baryons. We adopted the same criterion used by the main DES cosmological analysis (\citealt{y3-cosmicshear1}; \citealt*{y3-cosmicshear2}; \citealt{y3-3x2ptkp}). The criterion requires the peak of the marginalised two dimensional posterior of $\Omega_{\rm m}$ and $S_8 \equiv \sigma_8 (\Omega_{\rm m}/0.3)^{0.5}$ obtained by analysing the contaminated data vector to be within $0.3 \sigma$ of the values obtained with the uncontaminated one. We note that the baryonic model adopted by the main DES analyses to determine the scale cut follows the predictions from the OWLS `AGN' simulations \citep{Schaye2010,vanDaalen2011}. The baryonic feedback of the \texttt{CosmoGridV1} simulations, however, is slightly milder then the OWLS model. This difference is illustrated in Fig.~\ref{fig:effectofbaryons}, where we also show the impact of the OWLS AGN feedback on second moments, computed following the method in \citealt{moments2021}.

The level of contamination obtained using all the scales at our disposal is reported in Table~\ref{fig:contamination}, for a subset of individual summary statistics and for (some) of their combinations. Fortunately, none of the summary statistics exceed our predefined criteria for contamination, which confirms the robustness of our analysis against potential baryonic feedback processes. While it is true that the baryonic model of the \texttt{CosmoGridV1} simulations is milder than the OWLS model, these numbers are safely smaller than $0.3 \sigma$. {For second moments only, where we can compute the impact of the OWLS AGN feedback analytically, we also analysed a theory data vector `contaminated' with the OWLS AGN feedback, finding only a 0.1$\sigma$ shift with respect to dark-matter-only data vector.} In hindsight, we realize that we could have generated maps with higher resolution, even beyond \textsc{NSIDE} = 512 ($\sim 7$ arcmin). Such higher resolution would have allowed us to explore smaller scales, but it would have come with a considerable increase in computational cost, which we choose to defer to future research. 

\subsection{Impact of potential mismodelling of source clustering effects}

%

Source clustering refers to the angular distribution of source galaxies being not uniform, but rather being modulated by clustering due to galaxies tracing the underlying density field \citep{Schneider2002,Schmidt2009,Valageas2014,Krause2021,Gatti2023}. This effect causes the galaxy number density to be correlated with the target lensing signal: since we expect a larger lensing signal along overdense lines-of-sight, we preferentially sample the shear field where its value is larger. 

For estimators based on pixelized shear maps, this has two effects \citep{Gatti2023}:
\begin{itemize}
    \item the average noise-free lensing signal is modulated by a different effective redshift distribution;
    \item the shape noise in every pixel is correlated with the lensing signal.
\end{itemize}

The first effect is generally small. The second effect can be large for non-Gaussian statistics whenever the estimators used involve correlation between the lensing signal and even moments of the noise (e.g. in the case of third moments). Both effects impact mostly small scales. In this work, source clustering in our simulations was forward modelled following the prescription presented in \cite{Gatti2023} (see Eq.~\ref{eq:sc_pixel}). This implementation assumes a linear galaxy-matter bias for our sample. Furthermore, for simplicity, we also chose not to marginalise over such a bias, instead fixing its value to unity. We took some precautions to minimise the effect of source clustering in case our source clustering model does not faithfully reproduce the effects on data (which might happen, for instance, if the galaxy-matter bias of the source was different from unity). In \cite{Gatti2023}, the authors pointed out that for third moments the largest effect due to source clustering is related to the spurious signal-noise correlations, and that this can be removed completely by subtracting from the third moments estimators specific moments involving combinations of the observed noisy maps and noise-only maps (see \S \ref{sect:2nd3rd}). For the other statistics used in this work, we tested that source clustering effects are most noticeable for WPH $S01$ and WPH $C01$, and negligible for the other statistics. For WPH $S01$ and WPH $C01$, therefore, we adopted a noise-subtraction procedure similar to the one applied to third moments  (see \S \ref{sect:WPH}, \ref{sect:ST}), which we  empirically found to reduce the impact of source clustering on the measurements. 


In order to test the impact of any potential mismodelling of source clustering effects on our results, we analysed two sets of maps generated assuming a galaxy-matter bias $b=0.5$ or $b=1.5$ instead of unity. We verified that in none of our combination of summary statistics did the bias in the $S_8$-$\Omega_{\rm m}$ plane exceed 0.10$\sigma$. This means that the impact on cosmological parameters is safely negligible and that our modelling of source clustering is sufficiently accurate that small scales need not be removed from our analysis.

\subsection{Impact of additive biases due to PSF errors}\label{sect:PSF_modelling_errors}

We assess here the degree of contamination in our data vector resulting from the inclusion of additive biases associated with the misestimation of the Point Spread Function (PSF). The misestimation of the PSF can introduce additional biases in the measured shapes of galaxies, leading to deviations from their true values:
\begin{equation}
\vecgest=\vecg +\delta \vest^{\textrm{sys}}_{\textrm{PSF}}+\delta\vest^{\textrm{noise}}.
\end{equation}

To quantify these unwanted contributions, we can employ a model that accounts for the errors in PSF modelling and use a catalogue of `reserved' stars. These reserved stars are not used in training the PSF model and serve as a reference to characterize the spurious effects accurately. We follow \cite{Jarvis2016} and \cite*{y3-shapecatalog} by assuming that
\begin{equation}
\label{eq:new}
\delta \vest^{\textrm{sys}}_{\textrm{PSF}}=\alpha \vest_{\rm model}+\beta\left(\vestt- \vest_{\rm model}\right)+\eta\left(\vestt\frac{T_{\textrm{\rm *}}-T_{\rm model}}{T_{\rm *}}\right),
\end{equation}
where $\alpha,$ $\beta$, and $\eta$ are coefficients estimated from data, $\vestt$ is the PSF ellipticity measured directly using the reserved stars catalogue, $T_{\rm model}$ is the modelled PSF size, and $T_{\rm *}$ is the PSF size measured from the reserved stars catalogue. The coefficients $\alpha,$ $\beta$, and $\eta$ for the DES Y3 shape catalogue for the four tomographic bins are provided in \cite{y3-cosmicshear1}. 

We use an empirical method to estimate the contribution  of PSF additive biases to the summary statistics used in this work. We first created maps of $\vest_{\rm model}$, \vestt, and $\vestt \frac{T_{\textrm{\rm *}}-T_{\rm model}}{T_{\rm *}}$ from the reserved stars catalogue. Using the estimated values for $\alpha,$ $\beta$, and $\eta$, we then created maps of $\delta \vest^{\textrm{sys}}_{\textrm{model}}$, one for each tomographic bin. We added these systematic maps to a set of simulated maps at the fiducial cosmology, and proceeded to compute the summary statistics and to analyse the measurement with our LFI pipeline. We repeated the same procedure on maps with no PSF additive biases, and compared the two analyses at the level of the constraints in the  $S_8$-$\Omega_{\rm m}$ plane. We verified that in none of our combinations of summary statistics did the bias in the $S_8$-$\Omega_{\rm m}$ plane exceed 0.10$\sigma$, indicating that PSF modelling errors are negligible for the range of scales used in this work.

\section{End-to-end tests on simulations}\label{sect:end-to-end}

\begin{table*}
\caption {Constraints on various parameters for different summary statistics and their combinations. All the other parameters are marginalised over (see Table~\ref{parameter} for a list of parameters and their priors). For each parameter we report the 68 per cent confidence interval; numbers in parentheses refer to the percentage gain (or loss) with respect to the constraints from the second moments. Note that the improvement on the FOM in the last column is the most meaningful metric of a method's statistical power.}
\centering
\begin{tabular}{c|c|c|c|c|c|c}
 \hline
\textbf{Summary Statistic(s)}& \textbf{$\sigma(S_8)$} &  $\sigma(\sigma_8)$& $\sigma(\Omega_{\rm m})$  & $\sigma(w)$  & $\sigma(A_{\rm IA})$ & $ {\rm FoM 
 (S_8,\Omega_{\rm m})}$ \\
 & [x100]& [x100] & [x100]  & [x10] & [x10] & - \\
 \hline
2nd moments & 2.7 &  5.3  &  3.4   &  1.3  &  4.4  &  904 \\
2nd + 3rd moments & 2.6({\color{OliveGreen} + 3\%})& 5.0({\color{OliveGreen} + 6\%})& 3.4({\color{red} -0\%})& 1.3({\color{red} -3\%})& 4.2({\color{OliveGreen} + 5\%})& 1035({\color{OliveGreen} +15\%})\\
2nd moments + ST & 2.7({\color{OliveGreen} + 2\%})& 4.3({\color{OliveGreen} +19\%})& 3.0({\color{OliveGreen} +12\%})& 1.2({\color{OliveGreen} +11\%})& 4.4({\color{OliveGreen} + 0\%})& 1245({\color{OliveGreen} +38\%})\\
2nd moments + WPH & 2.4({\color{OliveGreen} +11\%})& 4.4({\color{OliveGreen} +18\%})& 2.9({\color{OliveGreen} +15\%})& 1.1({\color{OliveGreen} +15\%})& 3.9({\color{OliveGreen} +10\%})& 1385({\color{OliveGreen} +53\%})\\
2nd moments + ST + WPH & 2.0({\color{OliveGreen} +25\%})& 4.4({\color{OliveGreen} +18\%})& 2.9({\color{OliveGreen} +15\%})& 1.2({\color{OliveGreen} + 9\%})& 3.6({\color{OliveGreen} +17\%})& 1684({\color{OliveGreen} +86\%})\\
2nd + 3rd moments + ST + WPH & 2.0({\color{OliveGreen} +25\%})& 3.9({\color{OliveGreen} +26\%})& 2.9({\color{OliveGreen} +14\%})& 1.2({\color{OliveGreen} +12\%})& 3.6({\color{OliveGreen} +17\%})& 1733({\color{OliveGreen} +92\%})\\
\hline
\end{tabular}
\label{fig:improv}
\end{table*}

\begin{figure*}
\begin{center}
\includegraphics[width=0.9\textwidth]{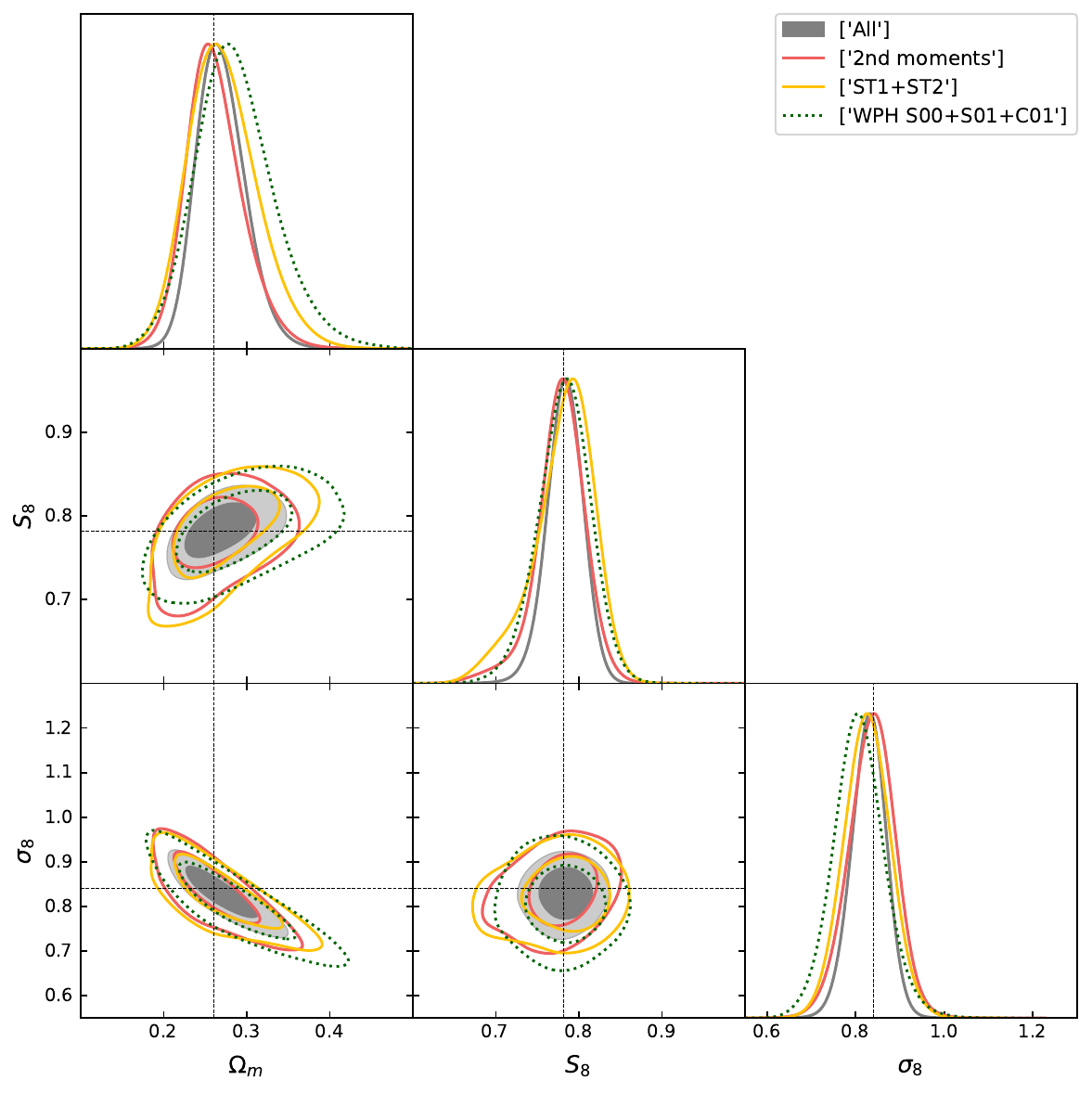}
\end{center}
\caption{Posterior distributions of the cosmological parameters $\Omega_{\rm m}$, $S_8$, and $\sigma_8$, for different summary statistics and their combinations, as measured in \texttt{CosmoGridV1} simulations. The dotted black lines indicate the values of the cosmological parameters in the simulations. The two-dimensional marginalised contours in these figures show the 68 per cent and 95 per cent confidence levels.}
\label{fig:all_}
\end{figure*}

\begin{figure}
\begin{center}
\includegraphics[width=0.36\textwidth]{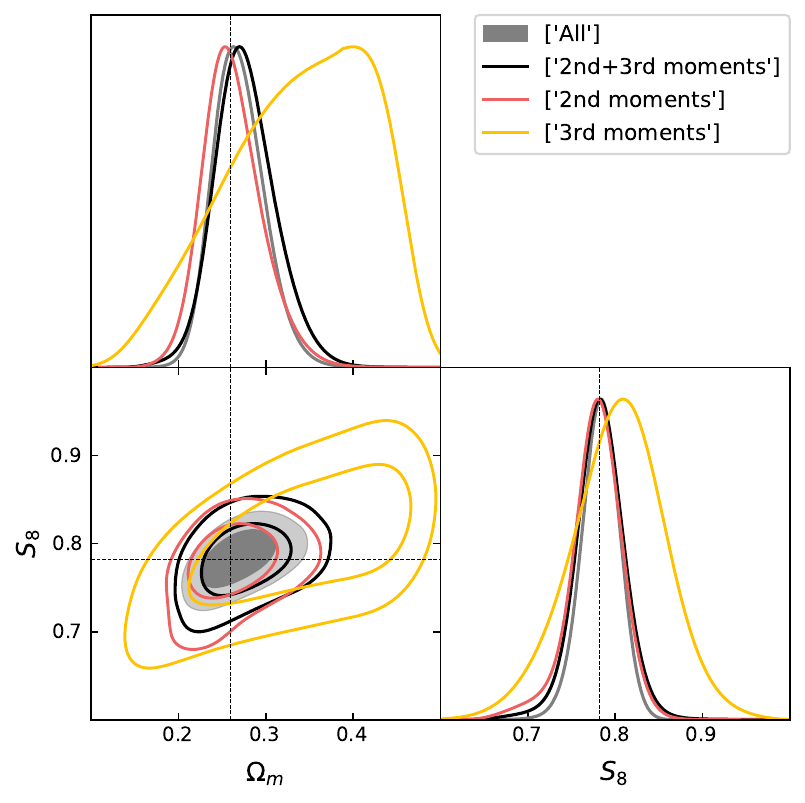}
\includegraphics[width=0.36\textwidth]{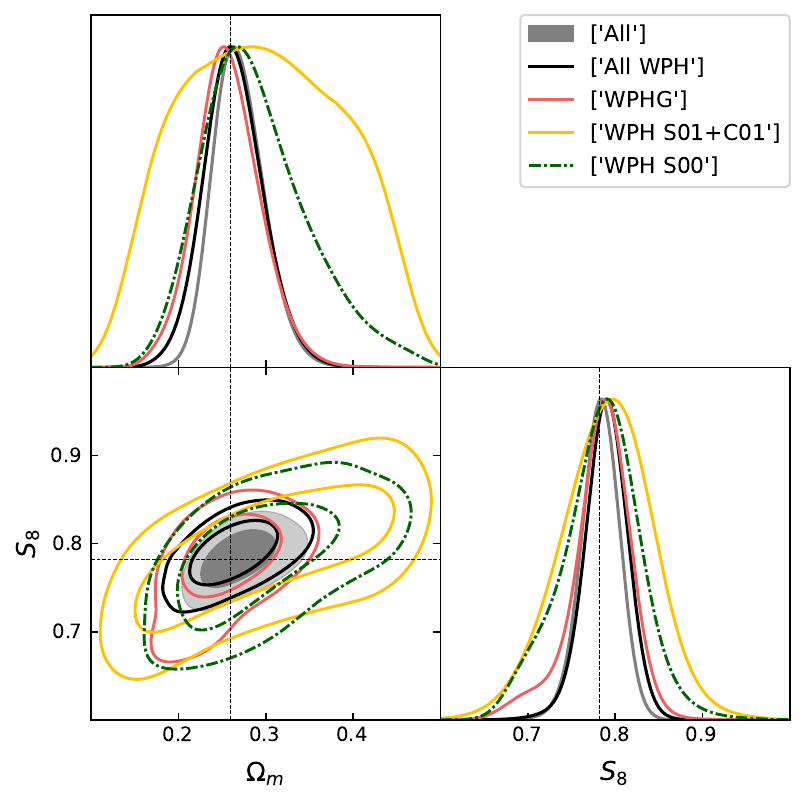}
\includegraphics[width=0.36\textwidth]{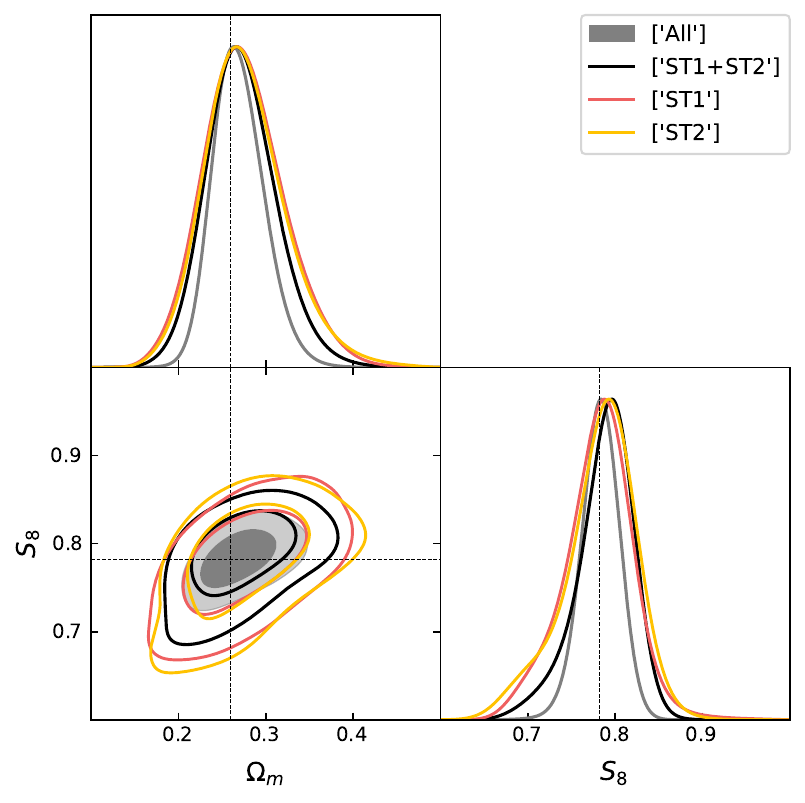}
\end{center}
\caption{Posterior distributions of the cosmological parameters $\Omega_{\rm m}$ and $S_8$, for different summary statistics and their combinations, as measured in \texttt{CosmoGridV1} simulations. The dotted black lines indicate the values of the cosmological parameters in the simulations. The two-dimensional marginalised contours in these figures show the 68 per cent and 95 per cent confidence levels.}
\label{fig:all_2}
\end{figure}

\begin{figure}
\begin{center}
\includegraphics[width=0.45\textwidth]{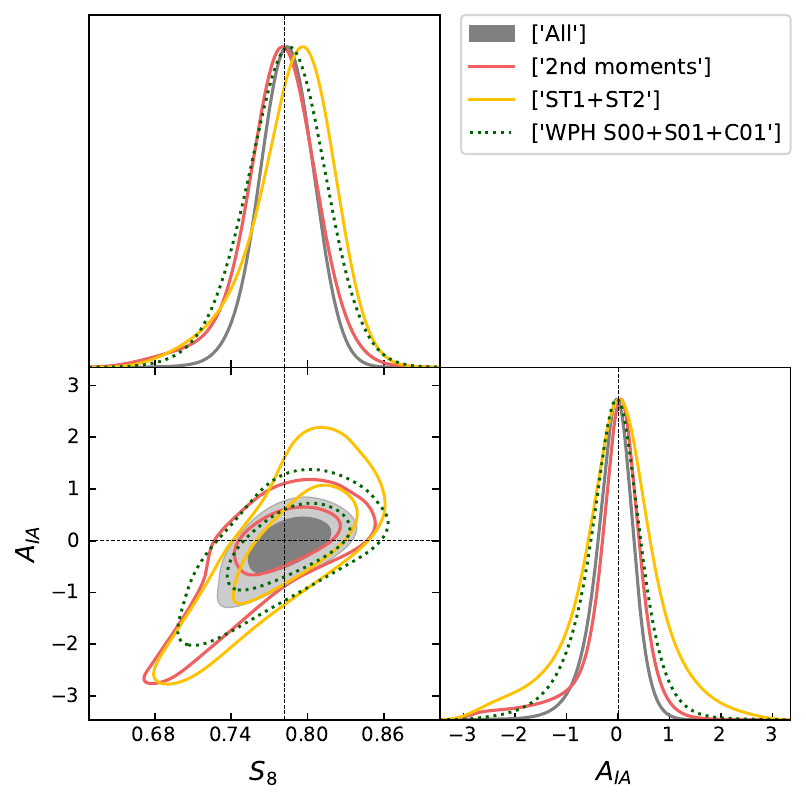}
\end{center}
\caption{Posterior distributions of the cosmological parameters $S_8$ and $A_{\rm IA}$, for different summary statistics and their combinations, as measured in \texttt{CosmoGridV1} simulations. The dotted black lines indicate the values of the cosmological parameters in the simulations. The two-dimensional marginalised contours in these figures show the 68 percent and 95 percent confidence levels. Note that there is significant improvement from using non-Gaussian statistics in the 95 percent confidence levels but less so in the 68 percent levels. }
\label{fig:all_IA}
\end{figure}

\begin{figure}
\begin{center}
\includegraphics[width=0.45\textwidth]{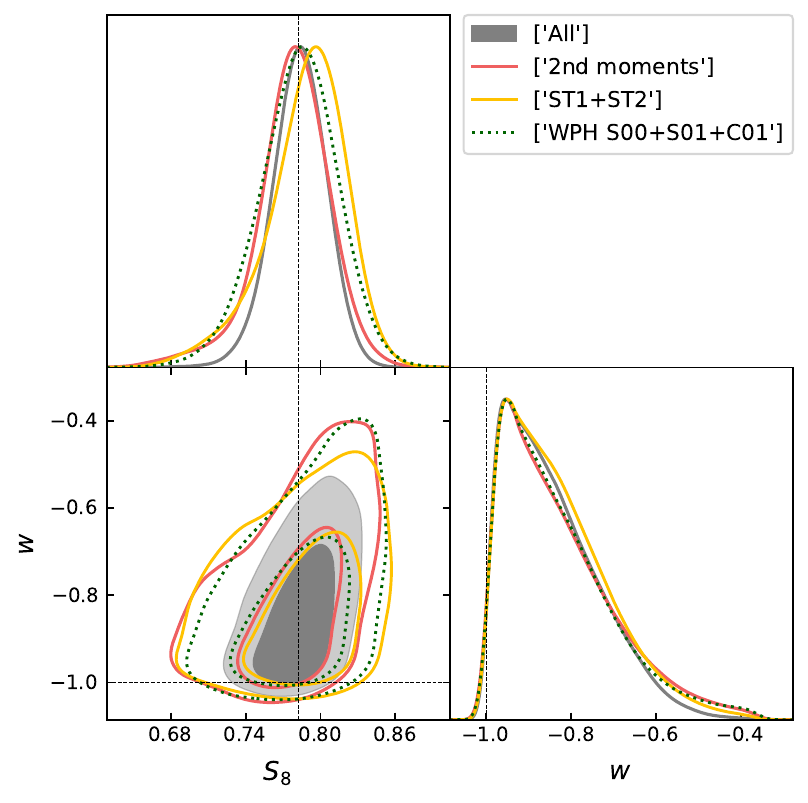}
\end{center}
\caption{Posterior distributions of the cosmological parameters $S_8$ and $w$, for different summary statistics and their combinations, as measured in \texttt{CosmoGridV1} simulations. The dotted black lines indicate the values of the cosmological parameters in the simulations. The two-dimensional marginalised contours in these figures show the 68 per cent and 95 per cent confidence levels.}
\label{fig:all_w}
\end{figure}


Having verified that all the scales used in our analysis are safe against a number of systematics, we next verify that we are able to recover the true cosmology of a set of simulations that have not been used to build our pipeline. To this end, we use 400 independent DES Y3 mock catalogues produced with the \texttt{CosmoGridV1} simulations. Each mock has the same cosmology; we further assume no intrinsic alignment, while for the other nuisance parameters (shear calibration and redshift uncertainties) we assume values at the centre of the priors. We measure all the summary statistics in the mocks, and then we average them, to reduce the impact of noise.

Our LFI analysis marginalises over seven cosmological parameters, assuming a $\nu w$CDM model; moreover, it marginalises over multiplicative shear bias (four parameters), intrinsic alignment (two parameters), and redshift distributions, as summarised in Table~\ref{parameter}. In addition to these parameters, we will also quote results in terms of the $S_8$ parameter, defined as
\begin{equation}
S_8 \equiv \sigma_8(\Omega_{\rm m}/0.3)^{\alpha} \,.
\end{equation} 
The value of $\alpha$ can be chosen so that $S_8$ best constrains the degeneracy between $\Omega_{\rm m}$ and $\sigma_8$. However, the summary statistics considered in this work have different directions and so there is no value of $\alpha$ that simultaneously optimises all. For sake of simplicity we adopt $\alpha=0.5$. We also quote a Figure-of-Merit (FoM), defined for $S_8$, $\Omega_{\rm m}$, and their
covariance:
\begin{equation}
{\rm FoM_{S_8,\Omega_{\rm m}}} = \left({\rm det} (C_{S_8,\Omega_{\rm m}})\right)^{-0.5}
\end{equation}

Fig.~\ref{fig:all_} shows the posteriors for $S_8$, $\Omega_{\rm m}$, and $\sigma_8$ for a combination of different summary statistics; posteriors for other summary statistics are shown in Fig.~\ref{fig:all_2} for $S_8$ and $\Omega_{\rm m}$. In Fig.~\ref{fig:all_}, `All' means that all the summary statistics are combined, except for WPHG, as we found it does not add additional information compared to second moments alone. For this reason we also chose to always use second moments as a default Gaussian statistic when combining with other non-Gaussian probes. Individual parameter constraints, together with the $\rm FoM_{S_8,\Omega_{\rm m}}$, are reported in Table~\ref{fig:improv}.


From Figs.~\ref{fig:all_} and \ref{fig:all_2} it can be noted that non-Gaussian statistics such as third moments and WPH S01 and C01 are characterised by a slightly different degeneracy tilt in the $\sigma_8$-$\Omega_{\rm m}$ plane compared to second moments. This distinction also becomes apparent in the $S_8$-$\Omega_{\rm m}$ plane, as the posteriors deviate from alignment with the $S_8$ axis. For other non-Gaussian statistics, such as ST1, ST2, or WPH S00, this is less evident, and is probably due to their being highly correlated with the second moments.

When all the summary statistics are combined, the gain in terms of constraining power over the standard Gaussian statistics (either second moments or WPHG) is substantial: the constraints on $S_8$ improve by $\sim 25$ per cent, whereas the gain in terms of ${\rm FoM_{S_8,\Omega_{\rm m}}}$ is $\sim 90$ per cent, i.e. almost double. This level of improvement is expected, and is due to the additional non-Gaussian information probed by the non-Gaussian WPH moments, ST, and third moments, and the degeneracy breaking.

When looking at the individual probes, we find that the WPHG are slightly less constraining than second moments alone ($\sim 10 $ per cent less constraining on the FoM).  As they both probe the power spectrum of the maps, this indicates that the spacing between the wavelet filters used for the WPHG (where each filter scale is double the size of the one preceding it) is inferior to the spacing of the top hat filters used for the second moments filters (where we considered more intermediate scales). A similar results was also found by \cite{Zuercher2022} using simulations. This problem could be mitigated by also introducing additional scales for the wavelet filters; we leave this exploration to future works. We also find that ST1 and ST2, either individually or combined, are not as constraining as second moments ($\sim 10 $ per cent less constraining on the FoM when combined), despite appearing to be highly correlated (Fig.~\ref{fig:corr}), and despite being  characterised by a high signal-to-noise (Table~\ref{scalesetc}). We found that this is due to a non-optimal information extraction from cross-bins (Eq.~\ref{eq:cross_maps}); including the cross-maps in the data vector for ST improves the constraints only a small amount, whereas second moments or WPHG significantly improve their constraints when cross-bins are included in the data vector. This would suggest a need to explore alternative ways of incorporating cross-correlation information among diverse fields within the ST framework. Alternatively, this lends support to the idea of employing WPH, which naturally facilitates the correlation of distinct fields. 



Next, of the three categories of non-Gaussian statistics examined in this study, the strongest performance -- in terms of constraining power when combined with second moments -- is exhibited by WPH, with ST following, and third moments trailing. Nevertheless, the combination of all the different statistics continues to enhance the constraints, underscoring that each statistic delves into slightly distinct information.

We next look into the constraints for the other parameters varied in this analysis.

Fig.~\ref{fig:all_IA} shows the constraints on $S_8$ and $A_{\rm IA}$ (the amplitude of IA) for some of the summary statistics (and their combinations) considered here (see also Table~\ref{fig:improv}). The amplitude of IA is one of the other main parameters constrained by weak lensing probes \citep{Dacunha2022}. The posteriors recover the correct value ($A_{\rm IA}=0$); interestingly, whenever second moments are combined with any of the non-Gaussian statistic considered here, constraints on $A_{\rm IA}$ are improved, up to almost 20 per cent. The parameter $\eta_{\rm IA}$ (which controls the redshift evolution of the IA amplitude) is not very well constrained as, for $A_{\rm IA}=0$, any value of $\eta_{\rm IA}$  would provide an equally good fit. Recall that we used a slightly simpler IA model than the fiducial DES Y3 analysis: ours does not include tidal-torque terms (because our current pipeline lacks the capability to compute these terms). {It is possible that the enhancement in constraining power resulting from the incorporation of the non-Gaussian statistics of these extra IA terms might be different the ones obtained for $A_{\rm IA}$ and $\eta_{\rm IA}$; we defer this investigation to future work.}

Fig.~\ref{fig:all_w} shows the constraints on $S_8$ and $w$ (see also Table~\ref{fig:improv}). The \texttt{CosmoGridV1} simulations used here have been produced assuming a $\Lambda$CDM cosmology ($w=-1$):  correctly, Fig.~\ref{fig:all_w} shows the posteriors skewed towards the edge of the prior. Despite these posteriors being partially prior-dominated, the combination of different non-Gaussian statistics improves the constraints on $w$ with respect to second moments by roughly 10 per cent. We also analysed the posterior distributions of the four parameters describing the shear multiplicative biases ($m_i$), and four parameters  ($\Delta z_i$) describing the shift in the mean redshift of the $n(z)$. The $\Delta z_i$ have been estimated for each of the multiple $n(z)$ realisations that have been used to produce our simulated maps with respect to the fiducial DES Y3 $n(z)$ given by the mean of all these realisations. As for the priors on $\Delta z_i$, we assumed them be Gaussian with zero mean and standard deviations equal to the spread of the shifts. These parameters are usually dominated by their priors, and, typically, conventional Gaussian statistics struggle to improve over these prior constraints. Some recent studies have pointed out the potential of non-Gaussian statistics for self-calibration, as evidenced by their ability to enhance precision beyond prior limitations in such parameters \citep{Pyne2021}. Indeed, we already saw this effect for $A_{\rm IA}$.  The posteriors for $\Delta z_i$ and $m_i$, however, were basically the same as their priors; we noted only a small improvement for the $\Delta z_i$  corresponding to the second, third, and fourth bins by 5-10 per cent for the combination of all the summary statistics.

\section{Conclusions}

In this methodology paper, we have presented an end-to-end simulation-based cosmological analysis of a set of Gaussian and non-Gaussian weak lensing statistics using detailed mock catalogues of the first three years of data of the Dark Energy Survey. Our main goals are to show the constraining power of wavelet based non-Gaussian statistics and to validate  a simulation based inference framework for a broad class of statistics for lensing surveys. 

{We considered the following summary statistics of weak lensing mass maps:  1) second and third moments; 2) wavelet phase harmonics (WPH); 3) the scattering transform (ST). Second moments are Gaussian statistics, whereas third moments probe additional non-Gaussian information of the fields.  The WPH moments are second moments of smoothed weak lensing mass maps that have undergone a non-linear transformation, allowing for the exploration of the non-Gaussian features of the field. The ST coefficients are built through a series of smoothing and modulus operations applied to the input field, followed by an average. The WPH and ST are often linked to convolutional neural networks (CNNs) because the definition of the statistics bears similarities to the architecture of CNNs (but note the latter requires training data). They  capture both Gaussian and non-Gaussian features of the fields; however, being only first or second order in the input data, they are generally more robust to noise than higher order moments. Moreover, in our implementation of the WPH and ST, we considered maps smoothed by directional wavelets, whereas for moments we only considered isotropic top-hat filters.}




Our analysis is fully based on simulations. We produced 791 full-sky $N$-body simulations, spanning seven cosmological parameters assuming a $\nu w$CDM cosmology: 
 $\Omega_{\rm m}$, $\sigma_8$, $n_s$, $h_{100}$, $\Omega_{\rm b}$, $w$, $m_{\nu}$. Using the $N$-body full-sky simulations, we generated almost 13000 pseudo-independent DES Y3 weak lensing mock mass maps, which we used for our inference pipeline. Our mock mass maps implement realistic masks, noise variations, source clustering of the sources, and include the following astrophysical observational systematic effects: intrinsic alignments, shear calibration, and redshift calibration biases. Our analysis is tomographic, i.e. we forward model the four tomographic bins and maps into which the DES Y3 weak lensing sample is divided.

 We implemented a neural network compression of the summary statistics, and we estimated the parameter posteriors using a likelihood-free-inference (LFI) approach, with a combination of Gaussian Mixture Density Networks and Masked Autoregressive Flows to estimate the likelihood surface from our mocks. We extensively validated our pipeline, testing the size of the posteriors with a coverage probability test, and comparing the posterior obtained from the LFI pipeline against a theory-based and Gaussian likelihood approach for the special case of Gaussian statistics (i.e. the power spectrum of the maps).

 We tested that the scales used in this work were not affected by systematics not properly modelled in our simulations: namely, baryonic feedback effects, PSF modelling errors, and differences in the prescriptions used to model source clustering. Finally, we tested our pipeline on a set of independent simulations that have not been used in our training process, demonstrating we could recover the true values of the cosmological parameters of the simulation.

 Of the three combinations of `non-Gaussian statistic plus second moment' examined, WPH exhibits the strongest constraining power, followed by ST, and then third moments. The combination of all the different statistics continues to enhance the constraints, underscoring that each statistic delves into slightly distinct information. In particular, we found that when all the summary statistics are combined, the constraints on $S_8$, $\Omega_{\rm m}$, and on the Figure-Of-Merit ${\rm FoM_{S_8,\Omega_{\rm m}}}$ are improved by roughly 25 per cent, 15 per cent,  and 90 percent, respectively, over the constraints from second moments. Similar gains are found on $w$ ($\sim$15 percent), and on the amplitude of intrinsic alignment ($\sim$20 percent). 
 
 This work highlights the importance of analysing probes of higher order statistics to improve the cosmological constraints, and showcases the power of a full simulation-based framework to efficiently model and combine different non-Gaussian probes.  Here we targeted the analysis at the third year (Y3) data from the Dark Energy Survey (DES), but the methodological advances presented here are suitable for application to Stage IV surveys from Euclid, Rubin-LSST, and Roman, once any necessary additional validation is caried out on mock catalogues for each survey. In a companion paper (Gatti et al., in prep.) we present an application to the DES Year 3 data.

\section*{Acknowledgements}

Funding for the DES Projects has been provided by the U.S. Department of Energy, the U.S. National Science Foundation, the Ministry of Science and Education of Spain, 
the Science and Technology Facilities Council of the United Kingdom, the Higher Education Funding Council for England, the National Center for Supercomputing 
Applications at the University of Illinois at Urbana-Champaign, the Kavli Institute of Cosmological Physics at the University of Chicago, 
the Center for Cosmology and Astro-Particle Physics at the Ohio State University,
the Mitchell Institute for Fundamental Physics and Astronomy at Texas A\&M University, Financiadora de Estudos e Projetos, 
Funda{\c c}{\~a}o Carlos Chagas Filho de Amparo {\`a} Pesquisa do Estado do Rio de Janeiro, Conselho Nacional de Desenvolvimento Cient{\'i}fico e Tecnol{\'o}gico and 
the Minist{\'e}rio da Ci{\^e}ncia, Tecnologia e Inova{\c c}{\~a}o, the Deutsche Forschungsgemeinschaft and the Collaborating Institutions in the Dark Energy Survey. 

The Collaborating Institutions are Argonne National Laboratory, the University of California at Santa Cruz, the University of Cambridge, Centro de Investigaciones Energ{\'e}ticas, 
Medioambientales y Tecnol{\'o}gicas-Madrid, the University of Chicago, University College London, the DES-Brazil Consortium, the University of Edinburgh, 
the Eidgen{\"o}ssische Technische Hochschule (ETH) Z{\"u}rich, 
Fermi National Accelerator Laboratory, the University of Illinois at Urbana-Champaign, the Institut de Ci{\`e}ncies de l'Espai (IEEC/CSIC), 
the Institut de F{\'i}sica d'Altes Energies, Lawrence Berkeley National Laboratory, the Ludwig-Maximilians Universit{\"a}t M{\"u}nchen and the associated Excellence Cluster Universe, 
the University of Michigan, NSF's NOIRLab, the University of Nottingham, the Ohio State University, the University of Pennsylvania, the University of Portsmouth, 
SLAC National Accelerator Laboratory, Stanford University, the University of Sussex, Texas A\&M University, and the OzDES Membership Consortium.

Based in part on observations at Cerro Tololo Inter-American Observatory at NSF's NOIRLab (NOIRLab Prop. ID 2012B-0001; PI: J. Frieman), which is managed by the Association of Universities for Research in Astronomy (AURA) under a cooperative agreement with the National Science Foundation.

The DES data management system is supported by the National Science Foundation under Grant Numbers AST-1138766 and AST-1536171.
The DES participants from Spanish institutions are partially supported by MICINN under grants ESP2017-89838, PGC2018-094773, PGC2018-102021, SEV-2016-0588, SEV-2016-0597, and MDM-2015-0509, some of which include ERDF funds from the European Union. IFAE is partially funded by the CERCA program of the Generalitat de Catalunya.
Research leading to these results has received funding from the European Research
Council under the European Union's Seventh Framework Program (FP7/2007-2013) including ERC grant agreements 240672, 291329, and 306478.
We  acknowledge support from the Brazilian Instituto Nacional de Ci\^encia
e Tecnologia (INCT) do e-Universo (CNPq grant 465376/2014-2).

This manuscript has been authored by Fermi Research Alliance, LLC under Contract No. DE-AC02-07CH11359 with the U.S. Department of Energy, Office of Science, Office of High Energy Physics.

\bibliography{bibliography}
\bibliographystyle{mn2e_2author_arxiv_amp.bst}



\appendix

\section{Noise properties of the simulations}\label{sect:noise_terms}

In this Appendix, we conduct several sanity checks to evaluate the noise characteristics of our simulations. Specifically, we ensure that the noise properties of the simulations encompass those of the actual data. The noise properties of the simulations should be mildly cosmology dependent, due to source clustering effects (see \S~\ref{sect:map_making_procedure}). 

To perform this test, we consider the following statistics: moments (second, third, and fourth order) and cumulative distribution functions (CDFs). The CDFs \citep{Anbajagane2023,Banerjee2023} for a given field are defined as the fraction of circles that have an enclosed value of the field larger than a given threshold:
\begin{equation}
    \CDF{(\theta, k = P(\kappa_{\theta} > k))},
\end{equation}
where $k$ is the threshold. The CDFs can be formally shown to contain all volume integrals of higher-order functions \citep{Banerjee2023}. We measure the CDFs across ten smoothing scales, spaced logarithmically between 3.2 and 200 arcmin;  for each scale, we use five thresholds $k  \in$ $\left[ -20,-6,-2,0, 2 \right] \times 10^{-3}$.  

\begin{figure*}
\includegraphics[width=0.8\textwidth]{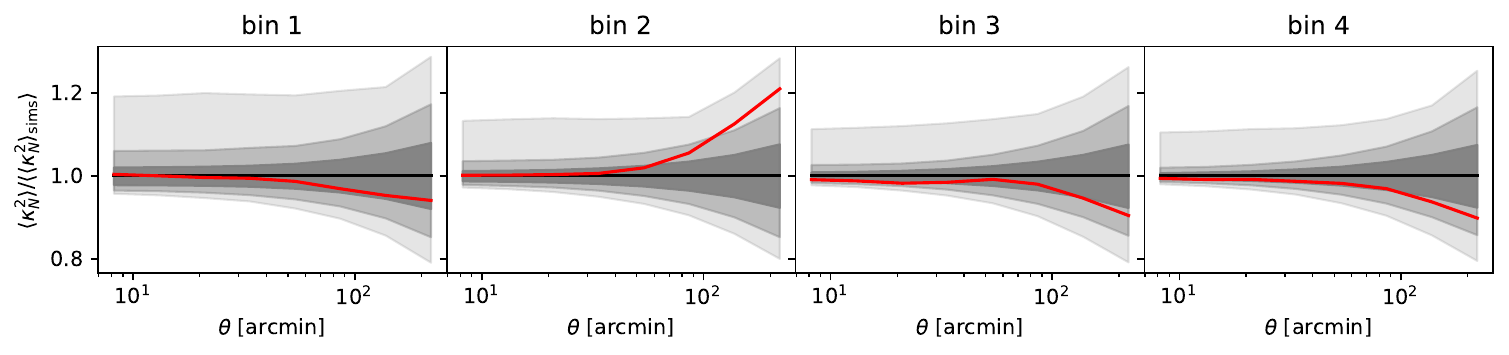}
\includegraphics[width=0.8\textwidth]{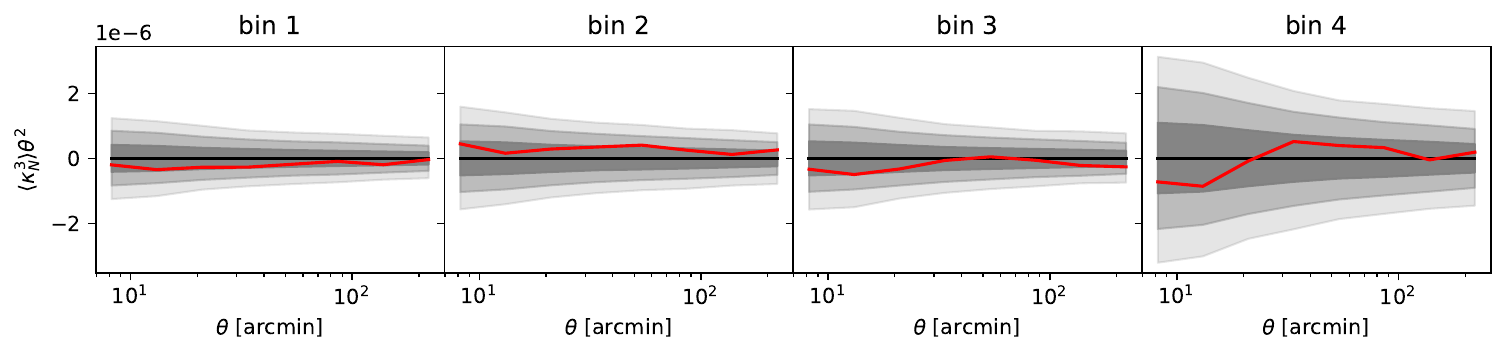}
\includegraphics[width=0.8\textwidth]{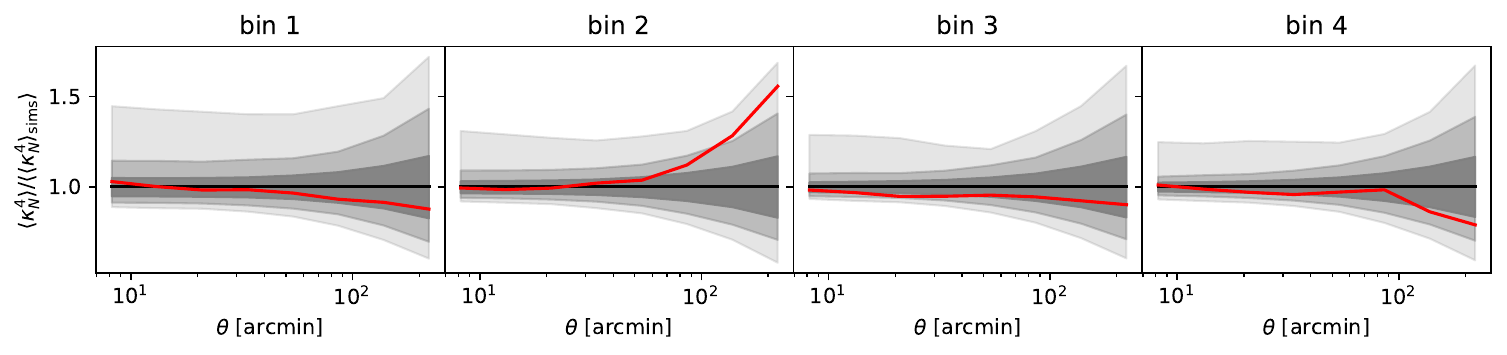}
\includegraphics[width=0.8\textwidth]{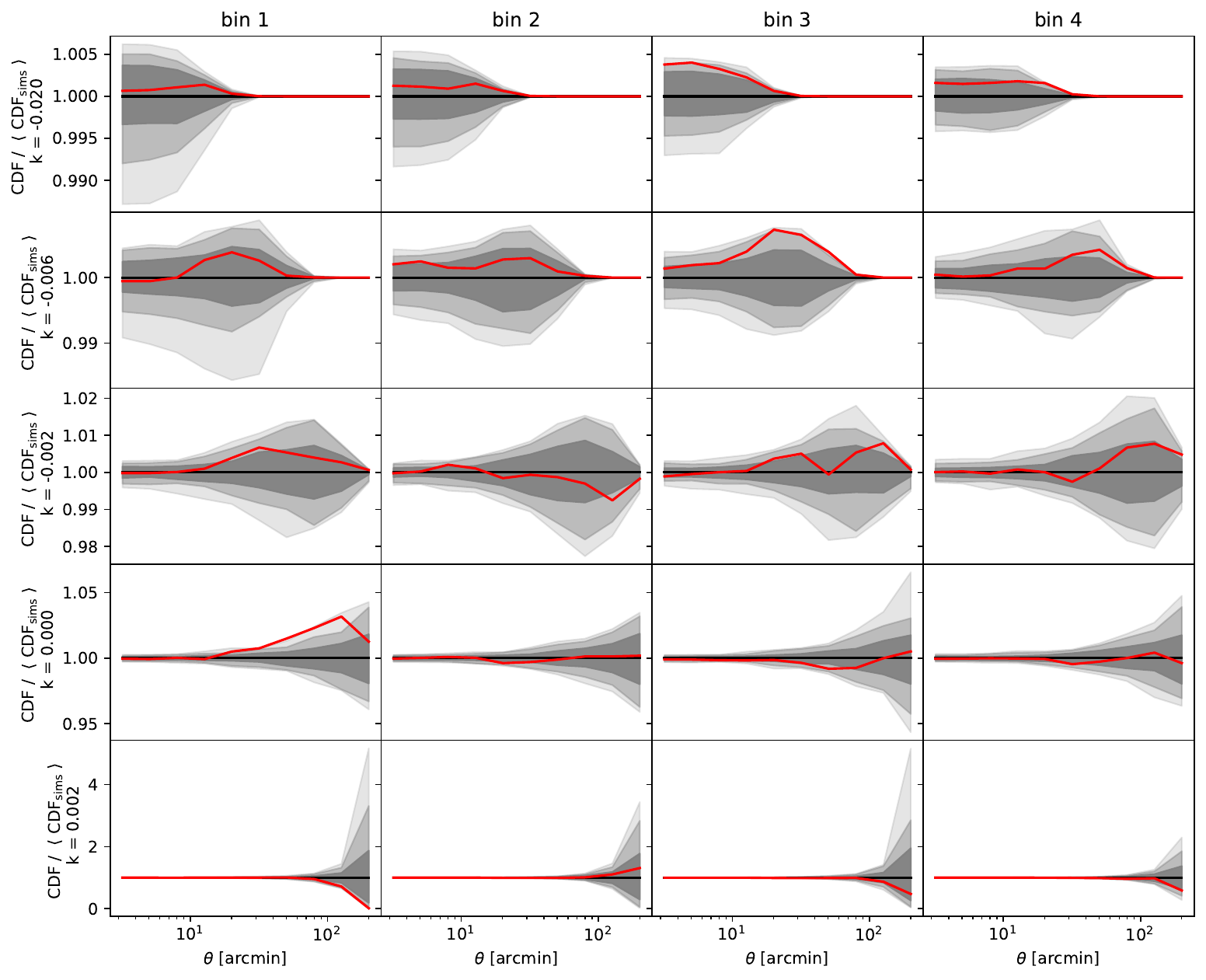}
\caption{Moments and CDFs of noise-only maps (grey shaded regions) in the Gower St simulations compared to the same quantities in data (red lines). The three different grey shaded regions encompass the 68, 95, and 99.5 percentiles spanned by the noise moments and CDFs in the simulations.}\label{fig:noise_terms}
\end{figure*}

In Fig.~\ref{fig:noise_terms} we compare a) moments and CDFs from data to b) moments and CDFs from simulations; there is a good match, indicating that the noise properties of our simulations reproduce well the noise of the data.

\section{Neural compression vs. MOPED compression}\label{sect:MOPED}

\begin{figure}
\includegraphics[width=0.45\textwidth]{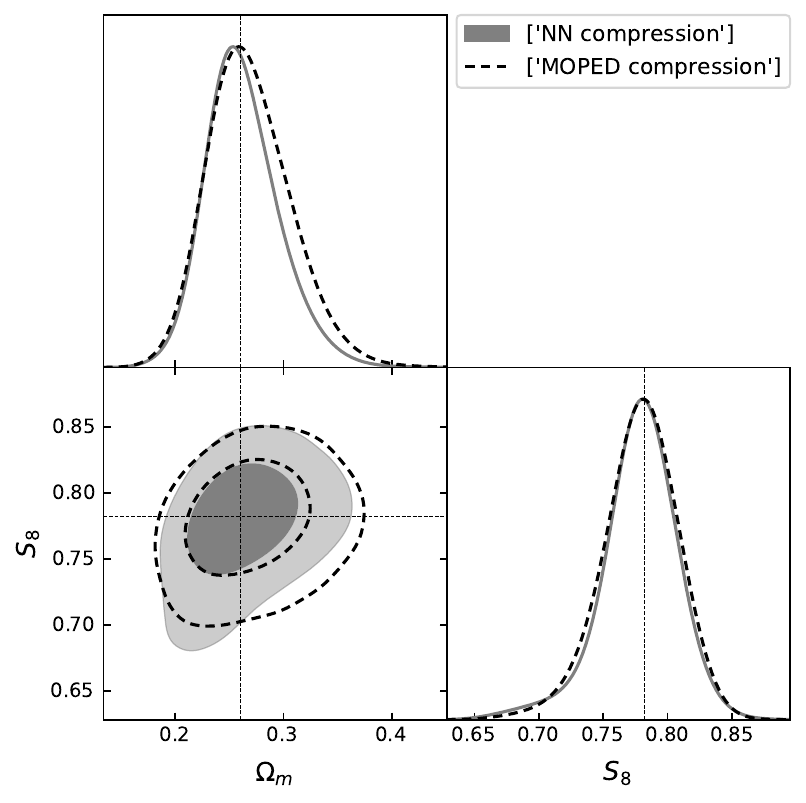}
\caption{Posterior distribution of the cosmological parameters $\Omega_{\rm m}$ and $S_8$ as measured in \texttt{CosmoGridV1} simulations. The two different posteriors have been obtained by analysing the second moments data vector compressed both with our fiducial neural network compression and with the alternative MOPED compression.} \label{fig:MOPED}
\end{figure}

In this work we opted for a neural compression scheme to compress our summary statistics. Other compression methods exist; the most notable is the MOPED algorithm \citep{Heavens2000}, which is lossless when the likelihood is Gaussian and the covariance matrix of the observables has a negligible dependence on the parameters. The neural network implemented in this work is in principle more powerful and general than the MOPED compression, as it does not make any assumption about the Gaussianity of the likelihood, nor about any dependence on the model parameters. Even if desired, we would not have been able to implement the MOPED compression for most of the statistics, as doing so would have required an estimate of the derivative of the model with respect to the parameters; such derivatives are available in closed form for analytical models, or via finite difference for observables where the model is estimated from simulations -- the Gower St simulations, however, do not allow us to estimate derivatives through finite differences. 

In this Appendix we compare the neural network compression with the MOPED compression for the only summary statistic for which we have an analytical model, i.e. the second moments. We also know that for second moments the likelihood should be fairly Gaussian, and the covariance should only weakly depend on parameters, so the MOPED compression should be close to lossless. We therefore compute the derivatives needed for the MOPED compression using the analytic model from \cite{G20}; for the covariance, we estimate it from the 400 measurements of the second moments in the \texttt{CosmoGridV1} simulations. Fig.~\ref{fig:MOPED} shows the posteriors obtained using our pipeline, compressing second moments either with the neural network or with the MOPED compression. Results are fairly similar, with the neural network compression delivering only slightly tighter contours.

\section{NDEs and parameters posterior}\label{sect:indivuals}

\begin{figure}
\includegraphics[width=0.45\textwidth]{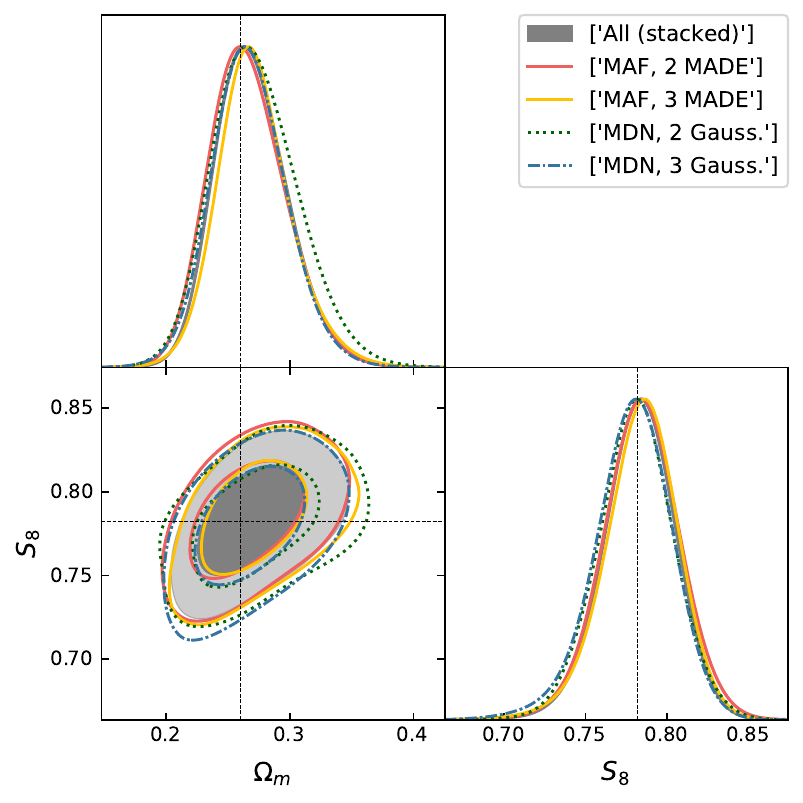}

\caption{Posterior distributions of the cosmological parameters $\Omega_{\rm m}$ and $\sigma_8$ for the combination of all the summary statistics considered in this work, as measured in \texttt{CosmoGridV1} simulations. We show the different posteriors as estimated by the different NDEs used in this work; we also show their stacked combination (the fiducial setup used in the other Figures of this paper). The dotted black lines indicate the values of the cosmological parameters in the simulations. The two-dimensional marginalised contours in these figures show the 68 per cent and 95 per cent confidence levels.}\label{fig:individual_posterior}
\end{figure}

In this work we used four different neural density estimators (NDEs) to estimate the posteriors. In particular, we used two different Gaussian Mixture Density Networks (MDNs) and two different Masked Autoencoders for Distribution Estimation (MADEs). Whenever we showed a posterior or reported the constraints on some parameters in this work, we always obtained these by stacking the four different NDEs. Assuming all the NDEs are flexible enough to describe our likelihood surface, they should all agree in the limit in which the number of simulations used for training becomes large. Fig.~\ref{fig:individual_posterior} shows the posteriors obtained by each individual NDE for our most constraining case (i.e. the combination of all summary statistics). We find that the 1 $\sigma$ constraints on $S_8$ do not vary more than 5 per cent across different NDEs. Although not shown here, we also repeated this test for all the other statistics (and combinations) considered in this work, and found differences below 5 per cent in all cases.

\section{Additional likelihood tests}\label{sect:likelihood_tests}
\begin{figure}
\includegraphics[width=0.45\textwidth]{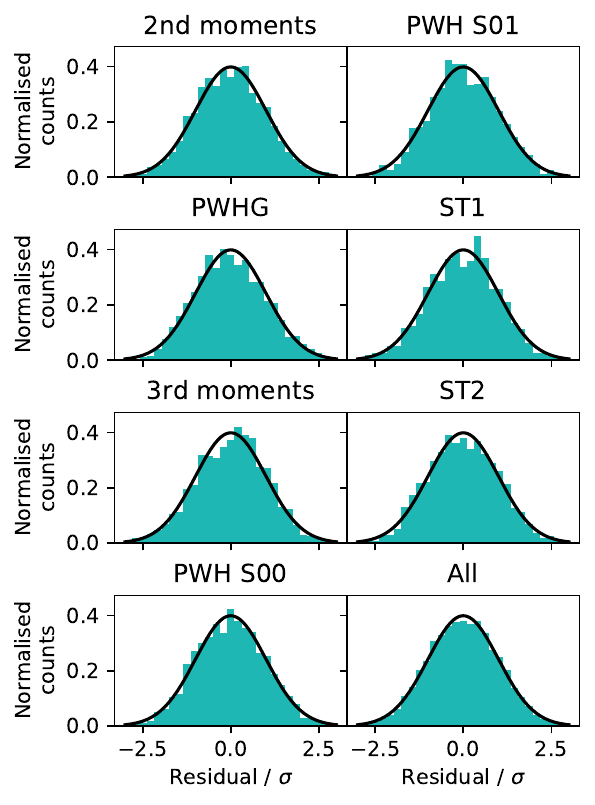}

\caption{Residuals of individual data points in units of their expected standard deviation for the compressed data vector of the \texttt{CosmoGridV1} simulations. We compare to a Gaussian with zero mean and unit standard deviation.}\label{fig:residuals}
\end{figure}

\begin{figure}
\includegraphics[width=0.45\textwidth]{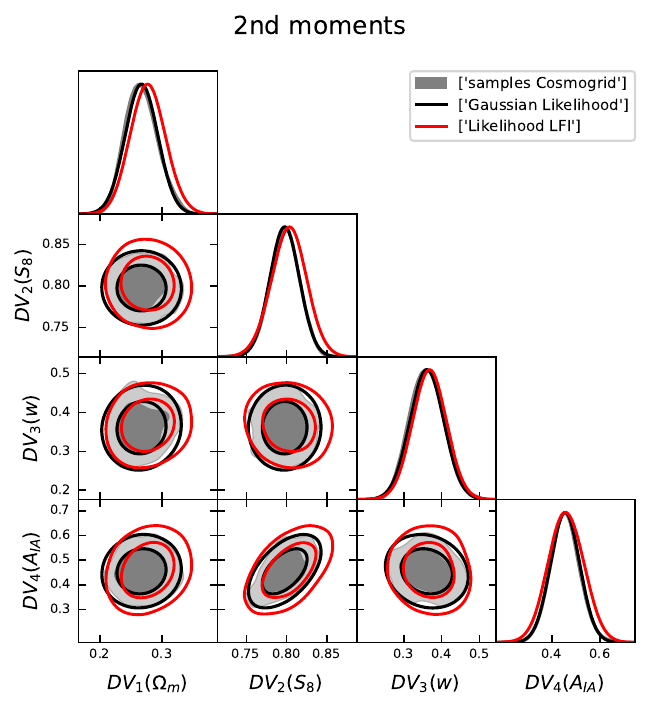}
\includegraphics[width=0.45\textwidth]{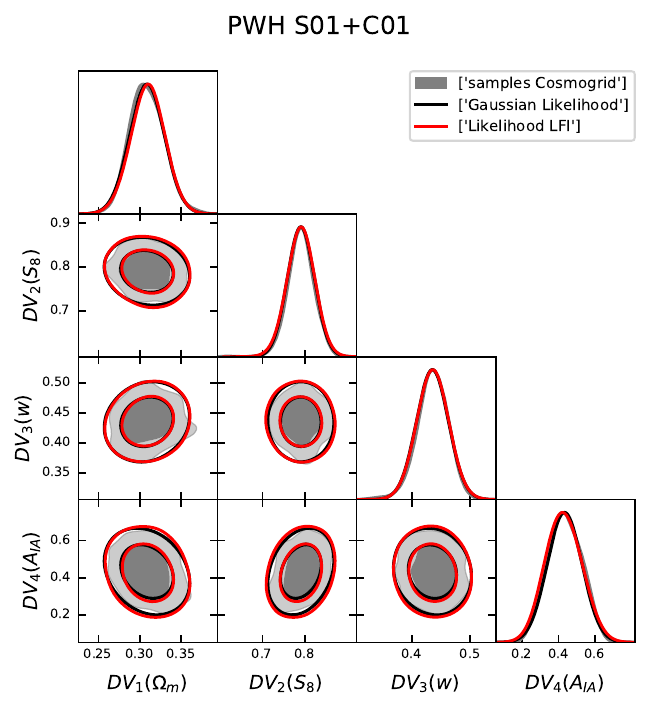}

\caption{Samples of compressed summary statistics from the \texttt{CosmoGridV1} simulations (grey), compared to samples drawn from the learned likelihood at the \texttt{CosmoGridV1} cosmology (red). We also compare to the distribution we would have obtained had we assumed a Gaussian likelihood (black). The top panel refers to second moments and the bottom panel refers to PWH S01+C01. }\label{likelihood_tests_dv}
\end{figure}

In this Appendix we perform extra tests on our estimated likelihoods using the \texttt{CosmoGridV1} simulations. First, for the 400 compressed data vectors at our disposal, we looked at the distribution of residuals for each entry of our data vector. This is shown in Fig.~\ref{fig:residuals}. The residuals are well described by a Gaussian, with no clear sign of strong deviations from Gaussianity. This is true also for the non-Gaussian statistics implemented in this work. As was already noted by \cite{G20}, this is partially thanks to the data compression algorithm, which helps to give the compressed data a more Gaussian distribution due to the central limit theorem \citep{Heavens2017}. 

As a second test, we sample from the likelihood estimated using our NDEs at the \texttt{CosmoGridV1} cosmology, and compare with the distribution of the compressed data vector measured in the \texttt{CosmoGridV1} simulations. In particular, we sample the likelihood at $\Omega_{\rm m} = 0.26$, $S_8 = 0.26 \sqrt{0.84/0.3}$, $w=-1$, and $A_{\rm IA} = 0$. For this test, we generated 400 new \texttt{CosmoGridV1} maps and we also marginalised over redshift uncertainties and multiplicative shear bias (in contrast to the \texttt{CosmoGridV1} maps used in the rest of the paper, where we fixed nuisance parameters to their mean values). Such a comparison is shown in Fig.~\ref{likelihood_tests_dv}, for the case of second moments and PWH S01+C01. Although not shown here, other summary statistics show a similar behaviour. The samples obtained from the NDEs match fairly well the distribution of compressed data vectors from the simulations, although for second moments they are slightly larger. This is expected: the likelihood estimated by the NDEs also marginalises over $\Omega_{\rm b}$, $n_s$, $h_{100}$, and neutrino mass, and we cannot fix them, because when training the NDEs we only made explicit  the dependence on  $\Omega_{\rm m}$, $S_8$, $w$, and $A_{\rm IA}$. The \texttt{CosmoGridV1} samples do not marginalise over these additional parameters, so their distribution might be slightly smaller than the one predicted from the NDEs.

In Fig.~\ref{likelihood_tests_dv} we also compare with the samples we would have obtained if we had assumed a Gaussian likelihood, estimating the mean and the covariance from the compressed \texttt{CosmoGridV1} measurements. These samples match very well the distribution of compressed measurements; this would not have been guaranteed had the likelihood been strongly non-Gaussian. Together with the residual tests (Fig.~\ref{fig:residuals}), this suggests that assuming a Gaussian likelihood for our compressed summary statistics could have been a reasonable option, at least at the \texttt{CosmoGridV1} cosmology. Of course, we cannot assume this generalises to other points in the parameter space, nor we could exclude \textit{a priori} any cosmological dependence of the covariance. We note that our NDEs have learned that the likelihood is Gaussian at this point in parameter space, as the NDEs did not have any prior knowledge concerning the form of the likelihood.


%


\label{lastpage}
\end{document}